\documentclass{article}
\usepackage{graphicx}
\usepackage{xcolor}
\usepackage{amsfonts}
\usepackage{amssymb}
\def\ligne#1{\hbox to\hsize{#1}}
\def\leurre{\noindent\leftskip0pt\small\baselineskip 10pt}
\newtheorem{thm}{\textbf{Theorem}}
\newtheorem{fig}{\textbf{Figure}}
\newtheorem{tab}{\textbf{Table}}

\author{Maurice {\sc Margenstern}}
\title{A weakly universal weighted cellular automaton in the heptagrid with 
6 states}
\begin{document}
\maketitle

\begin{abstract}
In this paper, we prove that there is a weakly universal weighted cellular 
automaton in the heptagrid, the tessellation $\{7,3\}$ of the hyperbolic plane, 
with 6 states. The present paper improves a previous result with 7 states deposited on 
arXiv:2301.1091v1, arXiv:2301.1091v2 and on arXiv:2301.1091v3.
\end{abstract}

\section{Introduction}~\label{intro}

    In the present paper see also \cite{mmarXiv23b}, the author considers weighted cellular 
automata in the heptagrid, the tessellation $\{7,3\}$ of the hyperbolic plane. He proves a theorem
about weak universality in that context, following a model used by the same author in many papers 
about cellular automata in hyperbolic spaces. By {\it weakly universal}, it is meant that the 
automaton is able to simulate a universal device starting from an infinite initial configuration. 
However, the initial configuration should not be arbitrary. It is the case as far as it is periodic 
outside a large enough circle as in previous mentioned papers, in fact it is periodic outside such 
a circle in two different directions as far as the simulated device is a two-registered machine. 
From a result by Minsky, \cite{minsky}, it is enough to simulate any Turing machine. In the 
heptagrid, the tessellation is based on a regular convex heptagon with the angle 
\hbox{$\displaystyle{{2\pi}\over3}$} between consecutive sides. The other tiles result by copies 
from that heptagon and, recursively, from the images by reflection in their sides. The heptagrid 
is defined and explained in Sub-section~\ref{sshepta}. In Sub-section~\ref{ssweight}, it is 
indicated what weighted cellular automata are. 
%In Section~\ref{sdodec} we extend the result to the dodecagrid,
%{\it i.e.} the tessellation $\{5,3,4\}$ of the hyperbolic $3D$-space which we also present in that 
%section.

\subsection{The heptagrid}\label{sshepta}

As already mentioned, the heptagrid is a tessellation of the hyperbolic plane whose signature is 
defined as \hbox{$\{7,3\}$}. In that signature, 7 is the number of sides of a tile, 3 is the 
number of edges which meet at a vertex. It also means that 3 is the number of tiles around a 
vertex. We call that tessellation, the {\bf heptagrid}. The signature also means that the tiles
are always copies of a {\bf heptagon}, a regular convex polygon of the hyperbolic plane with seven 
sides and with the angle \hbox{$\displaystyle{{2\pi}\over3}$} between consecutive sides as already
said.

The left-hand side part of Figure~\ref{fhepta} gives us a representation of the heptagrid. 
A {\bf path joining} $A$ to~$B$, 
where $A$ and $B$ are two tiles of the heptagrid, is a sequence \hbox{$\{T_i\}_{i\in\{0..n-1\}}$} 
of tiles such that $T_i$ and $T_{i+1}$ share a common side for \hbox{$0\leq i<n$-1} but $T_i$ and 
$T_j$ are not adjacent if \hbox{$\vert i-j\vert > 1$}. We say that $n$ is the length of the just 
mentioned path joining $A$ to~$B$. We call {\bf distance} from~$A$ to~$B$, denoted by dist$(A,B)$, 
the shortest length for the paths joining $A$ to~$B$. Clearly, dist$(A,B)=0$ if 
and only if $A=B$. Clearly too, that distance satisfies the triangular inequality. A circle of 
radius~$r$  around~$T$, a tile of the heptagrid, is the set of tiles of the heptagrid whose 
distance from~$T$ is~$r$. A {\bf disc} of radius~$r$ around~$T$ is the set of tiles in all circles 
of radius~$\rho$ around~$T$ with $\rho\leq r$. Figure~\ref{fhepta} also illustrates the main 
features which are used to navigate in the tiling. From the point $M$ of the figure, we draw 
two rays $u$ and $v$ which pass through the mid-points of the sides of tiles they are crossing,
a characteristic property. The tiles whose centre lies within the angle defined by the rays $u$
and $v$ constitute a {\bf sector}, by definition. The tile of the sector which is the closest to~$M$
is called its {\bf head}. The figure also illustrates the tree structure which allows us to 
navigate in the tiling. As illustrated by the right-hand side part of Figure~\ref{fhepta}, the 
heptagrid is the union of a central tile~$\tau$ and seven sectors whose heads are the neighbours 
of $\tau$. 

\newcount\compterel\compterel=1
\def\numerrel{\the\compterel\global \advance\compterel by 1}
%\vskip 10pt
\vtop{
\ligne{\hfill
\includegraphics[scale=0.4]{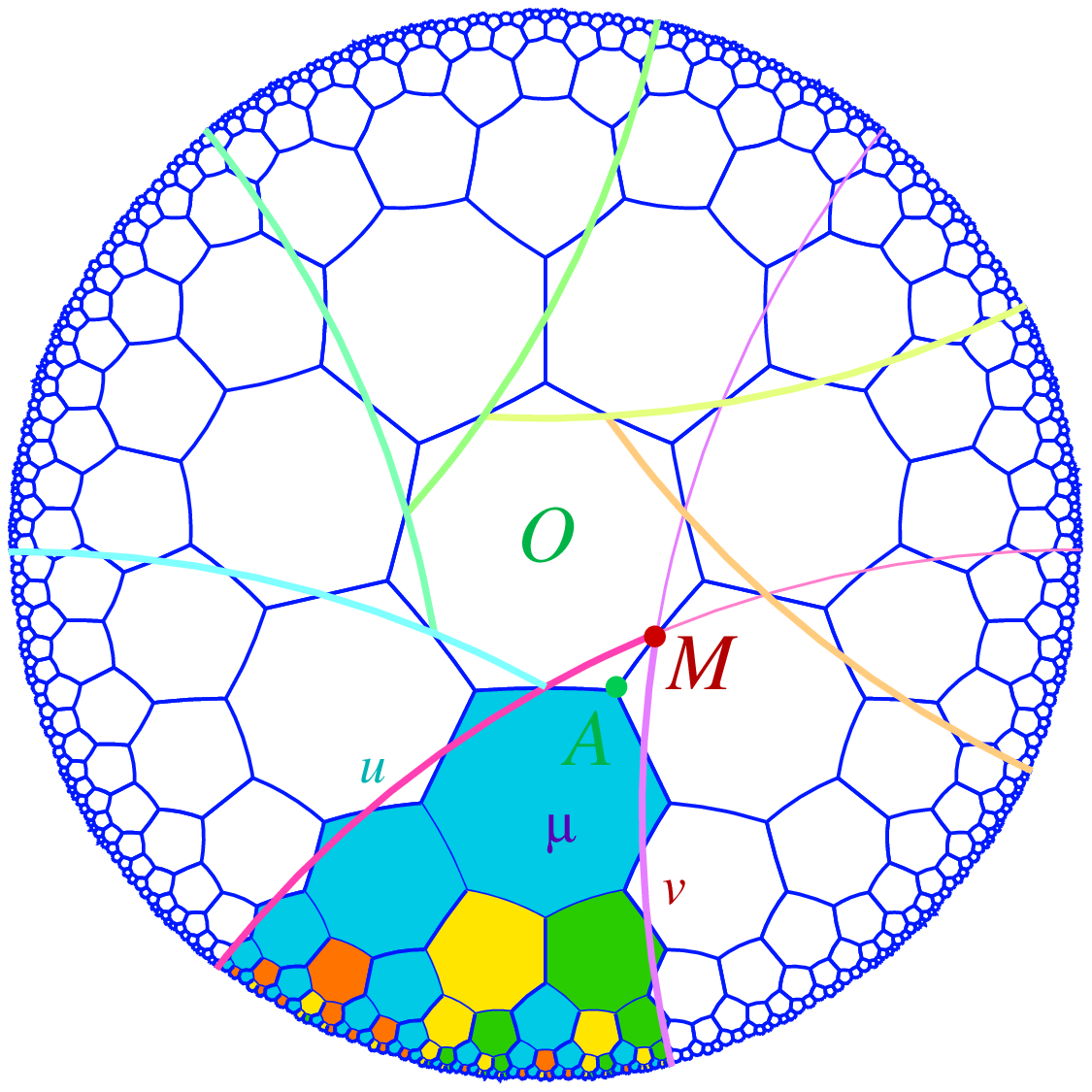}
\raise -5pt\hbox{\includegraphics[scale=1.65]{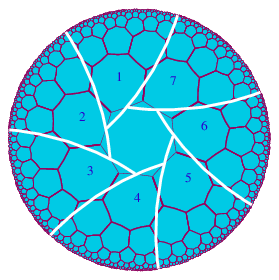}}
\hfill}
\vspace{-5pt}
\begin{fig}\label{fhepta}
\leurre
Representation of the heptagrid. To left, The rays $u$ and $v$ define a sector. Inside that sector, 
the tree structure. To right, the decomposition  of the heptagrid. Each sector is numbered in 
its {\bf head}. Accordingly, $\tau$ in the left-hand side picture is the head of sector~$4$, 
defined by $u$ and $v$; $\mu$'s coordinate is $(4,1)$. 
\end{fig}
}
\def\BB{\hbox{\bf B}}
\def\OO{\hbox{\bf O}}
\def\YY{\hbox{\bf Y}}
\def\GG{\hbox{\bf G}}
Indeed, if we denote the colours, blue, orange, yellow, orange and green by the letters \BB, \OO,
\YY{} and \GG, respectively, the tiling in a sector is defined by the rules:
\vskip 5pt
\ligne{\hfill$\vcenter{
\vtop{\leftskip 0pt\parindent 0pt\hsize=100pt
\ligne{\BB{} $\rightarrow$ \BB\OO}
\ligne{\OO{} $\rightarrow$ \BB\YY\OO}
\ligne{\YY{} $\rightarrow$ \BB\YY\GG}
\ligne{\GG{} $\rightarrow$ \BB\YY\GG}
}}$
\hfill(\numerrel)\hskip 15pt}
\vskip 5pt
In a rule of (1), we say that the tiles lying on the right hand-side of the arrow are {\bf produced}
by the rule from the tile lying on the left hand-side of the arrow. The tree $\mathcal T$ generated 
by those rules applied in the sector from the closest tile~$\tau$ to $M$ is called a 
{\bf Fibonacci tree} as far as there are $f_{2n+1}$ tiles which are at the distance $n$ from 
$\tau$, where $f_n$ is the sequence defined by 
\vskip 5pt
\ligne{\hfill$f_0=f_1=1$ and $f_{n+2} = f_{n+1}+f_n$ for $n\in\mathbb N$\hfill (\numerrel)
\hskip 15pt}
%   In previous papers about hyperbolic $3D$ cellular automata, I indicated a way to
%define the tiles within the Schlegel representation: i
%If we put a dodecahedron $\Delta_1$ 
%on a face~$i$ of~$\Delta$ we may represent~$\Delta_1$ by its projection on the plane 
%of the face~$i$. The problem of such a representation is that it should be partial in
%order to locate the dodecahedra. Their numbering is also possible as explained
%for instance in~\cite{mmbook1,smallbook}.  
\vskip 5pt
The navigation tool in a sector consists in numbering by~1 its closest tile~$\tau$ from~$M$ and 
numbering the others one by one from a level to the next one, a level in the sector being the set
of tiles at the same distance from~$\tau$, and on a level, from the leftmost tile to the rightmost
one. That numbering possesses interesting properties, we refer the reader 
to~\cite{mmbook1,smallbook}.

It is not difficult to see that if $A$ is a tile of~$\mathcal T$ at the distance~$n$ from~$\tau$
if and only if there is a sequence $\tau_i$ with \hbox{$i\in[0..n]$} such that $\tau_0=\tau$,
$\tau_n=A$ and for $0\leq i<n$, $\tau_{i+1}$ is a son of $\tau_i$, which means that $\tau_{i+1}$ is 
produced from $\tau_i$ by one of the rules of~$(1)$. Such a sequence is called the {\bf path from 
$\tau$ to $A$ in the sector}. In \cite{smallbook,mmbook1}, there are algorithms
allowing to compute the path from $\tau$ to $A$ from the number attached to~$A$ in the sector. 
%An infinite path from~$\tau_0$ is called a {\bf branch} of the Fibonacci tree or of the sector.
As can be seen on the right-hand side picture of Figure~\ref{fhepta}, the heptagrid can be split 
into a central tile~$O$ and into seven disjoint sectors whose union is the complement of~$O$ in the 
heptagrid. So that if a tile~$A$ of the heptagrid is distinct from~$O$, it can be given two 
numbers, $(s,n)$, $s$ in [1..7] defining in which sector it lies and $n$ being the number of 
the tile in its sector. Tile~$O$ is given number~0. The tile~$\mu$ in the left-hand side part of 
Figure~\ref{fhepta} is given the coordinate (4,1).
%it is the {\bf head} of sector~4.

\def\HH{$\mathcal H$}
   The cellular automaton we construct in Section~\ref{scenario} evolves in the
heptagrid. We shall use figures similar to Figure~\ref{fhepta} to illustrate key 
configurations while more complex ones will be illustrated by diagrams. 

   The grids of Figure~\ref{fhepta} can be reused without problem as far as there is no central tile
in the heptagrid. The representation we consider is based on the Poincar\'e's disc model of the
hyperbolic plane. There is no central point in the hyperbolic plane. We can see the disc model as 
a window over the hyperbolic plane, as if flying over that plane in an abstract spacecraft. The 
centre of the circle is the point on which are attention is focused while the circle itself is our 
horizon. Accordingly, the central tile is the tile which is central with respect to the area under 
our consideration. It is also the reason to number the central tile by~0.

    In a sector, the tile~$\mu$ which is closest to~$M$ is numbered by~1. It is the head of the 
sector which is sector~4{} in Figure~\ref{fhepta}, to left. The other tiles of 
that sector are numbered as already mentioned. Table~\ref{tnum} indicates the correspondence 
between the numbers of the tiles and their relations in the tree. A tile has also a {\bf level} 
which is its distance in the sector to the tile~$O$. In most representations, we will deal with 
levels~1,2 and~3 and sometimes, with level~4 too. Note that in Figure~\ref{fhepta} level~4 is
hardly visible.

Tiles which share a side are called {\bf neighbours} of each other. Sometime, we say that they can 
see each other. Accordingly, a path from a tile to another one~$A$ of the sector consists of a 
sequence of tiles of the sector which successively can see the previous one. When the path goes 
from~$\mu$ to~$A$ and when each tile of the path is the son of the previous one in the path we 
say that such a path belongs to the {\bf branch} in the tree which passes through~$A$, as far as 
for each tile~$A$ of the sector, there is a single branch of the tree passing through~$A$. 
Tile~$\mu$ is the head of the sector it defines, we also call it the {\bf root} of the tree which 
spans that sector. 
\vskip 5pt
\newdimen\ldima\ldima=35pt
\newdimen\ldimb\ldimb=28pt
\newdimen\ldimc\ldimc=17pt
\def\lignure #1 #2 #3 #4 #5 #6{%
\ligne{%
\hbox to \ldima{\hfill#1\hfill}\hbox to \ldimc{\hfill#2\hfill}\hbox to \ldimc{\hfill#3}
\hbox to \ldimb{\hfill#4}\hbox to \ldimb{\hfill#5} 
\hbox to \ldimb{\hfill#6\hfill}
\hfill}
\hfill}
\ligne{\hfill
\vtop{\leftskip 0pt\parindent 0pt\hsize=340pt
\begin{tab}\label{tnum}
\leurre	
Table of correspondence between numbers and their relations in the tree.
Nodes of colour \BB{} have two sons, while nodes of colours \OO, \YY{} and \GG{} have three of them.
Also, $\ell$-s., $m$-s. and $r$-s. mean left-hand side son, middle one and right-hand side one
respectively. By 'c.' we mean the colour of the node.
\end{tab}
\ligne{\hfill
\vtop{\leftskip 0pt\parindent 0pt\hsize=160pt
\lignure  {level} {node} {c.} {$\ell$-s.} {$m$-s.} {$r$-s.}
\lignure  0 1 G 2 3 4
\lignure  1 2 B 5 6 {}
\lignure  {} 3 Y 7 8 9
\lignure  {} 4 G {10} {11} {12}
\lignure  2 5 B {13} {14} {}
\lignure  {} 6 O {15} {16} {17}
\lignure  {} 7 B {18} {19} {}
\lignure  {} 8 Y {20} {21} {22}
\lignure  {} 9 G {23} {24} {25}
\lignure  {} {10} B {26} {27} {}
\lignure  {} {11} Y {28} {29} {30}
\lignure  {} {12} G {31} {32} {33}
}
\hfill
\vtop{\leftskip 0pt\parindent 0pt\hsize=160pt
\lignure  {level} {node} {c.} {$\ell$-s.} {$m$-s.} {$r$-s.}
\lignure  3 {13} B {34} {35} {}
\lignure  {} {14} O {36} {37} {38}
\lignure  {} {15} B {39} {40} {}
\lignure  {} {16} Y {41} {42} {43}
\lignure  {} {17} G {44} {45} {46}
\lignure  {} {18} B {47} {48} {}
\lignure  {} {19} O {49} {50} {51}
\lignure  {} {20} B {52} {53} {}
\lignure  {} {21} Y {54} {55} {56}
\lignure  {} {22} G {57} {58} {59}
\lignure  {} {23} B {60} {61} {}
\lignure  {} {24} Y {62} {63} {64}
\lignure  {} {25} G {65} {66} {67}
\lignure  {} {26} B {68} {69} {}
\lignure  {} {27} O {70} {71} {72}
\lignure  {} {28} B {73} {74} {}
\lignure  {} {29} Y {75} {76} {77}
\lignure  {} {30} G {78} {79} {80}
\lignure  {} {31} B {81} {82} {}
\lignure  {} {32} Y {83} {84} {85}
\lignure  {} {33} G {86} {87} {88}
}
\hfill}
}	
\hfill}
\vskip 5pt

\subsection{Weighted cellular automata}\label{ssweight}

Cellular automata are a model of massive parallelism. The base of a cellular automaton is a cell. 
The set of cells is supposed to be homogeneous in several aspects: the neighbours of each cell 
consists of a subset which has the same structure; the cell changes its state at each tip of a 
discrete clock according to the states of its neighbours and to its own state. The change is 
dictated by a finite automaton which is the same for each cell. A regular tiling is an appropriate 
space for implementing cellular automata: a cell is the combination of a tile together with the 
finite automaton ruling the change of states. The tile is called the {\bf support} of the cell. 
Let $T$ be a tile and let $N(T)$ be the set of its neighbours. By regular, we mean that the 
number of elements of $N(T)$ is the same for any~$T$. The heptagrid satisfies that 
requirement. Moreover, there is an algorithm to locate the tiles which is linear in time 
in the size of the code attached to each tile, see \cite{mmbook2} for instance. 
From now on we indifferently say tile or cell for a heptagon of the heptagrid, confusing the cell 
with its support. Sometimes, we also refer to a tile or to a cell by the number of its support in 
the sector it is in the figure illustrating the situation in which the cell is considered.

    The way the automaton manages the change of states is defined by what is called a 
{\bf transition function} which is often implemented as a table.  That function is called the 
{\bf program} of the automaton and we shall organise it in a {\bf table} which will be displayed 
in Section~\ref{stable}. In the present paper, we append a constraint on the transition function: 
states are affected with {\bf weights}, which are non negative integers. Consider a cell~$c$ 
together with its neighbours $c_i$, with \hbox{$i\in\{1..7\}$}. Let $s_i$ be the state of 
neighbour~$i$, which is $c_i$, and let $w_i$ be the weight of $s_i$. Call the {\bf neighbourhood 
weight} of $c$ the sum $s = \displaystyle{\sum_{i=1}^7w_i}$. In our paper, the new state of~$c$ is 
defined by its current stated together with $s$, its neighbourhood weight. In particular, the 
transition from the current state to the new one does not depend on the positions of those 
neighbours but on the sum of their weights only. That entails another constraint on the program as 
far as our cellular automaton is deterministic: a current state with a given neighbourhood weight 
give rise to single new state. A cellular automaton whose transitions obey such a 
constraint is called {\bf weighted}.

In Section~\ref{stable}, we define tables on which the transitions of the cellular automaton are
based. The tables have two entries: the current state of the cell and the neighbourhood weight for
that cell. The tables are gathered in Table~\ref{t_small} which provides us with the new state 
of the cell.

The alphabet $\mathbb A$ of the automaton attached to each cell is the set of the possible states 
taken by the cell. To each state, we attach a {\bf weight}, as already indicated. 
The function giving its weight for each state can be represented by a sequence of those weights,
giving an order on the states. So that writing \hbox{${\mathbb A} = \{e_0,...,e_n\}$}, 
the weights are \hbox{$\{w_0,...,w_n\}$}. As far as we already defined the neighbourhood weight 
of a cell, we presently turn to the construction of the table.

   Now that the global setting is given, we shall proceed as follows: 
Section~\ref{scenario} indicates the main lines of the implementation which is precisely
described in Subsection~\ref{newrailway}. At last, Section~\ref{stable} gives us the tables ruling
the transitions followed by the automaton. Subsection~\ref{newrailway} also contain a few figures 
which illustrate the application of the function. Those figures were established from pieces of 
figures drawn by a computer program which applied the transition function of the automaton to an 
appropriate window in each of the configurations described in Subsection~\ref{newrailway}. The 
computer program also computed the neighbourhood weight of a cell. It established the tables
displayed in Section~\ref{stable}.

   That allows us to prove the following property:

\begin{thm}\label{letheo}
There is a weakly universal weighted cellular automaton in the heptagrid which has six states. 
The highest weight of the states is $34$ and the maximal neighbourhood weight is
$156$. The table contains $137$ entries.
\end{thm}

%   The theorem was proved with the help of a computer program. 
The states and their weights used in the simulation proving the theorem are the following ones:
\newdimen\epaiss\epaiss=30pt
\vskip 5pt
\ligne{\hfill
\vtop{\leftskip 0pt\parindent 0pt\hsize=300pt
\ligne{\hbox to 40pt{states\hfill} 
\tt
\hbox to \epaiss{\hfill W\hfill} \hbox to \epaiss{\hfill Y\hfill} \hbox to \epaiss{\hfill B\hfill}
\hbox to \epaiss{\hfill R\hfill} \hbox to \epaiss{\hfill M\hfill} 
\hbox to \epaiss{\hfill V\hfill} %\hbox to \epaiss{\hfill V\hfill}
\hfill
}
\vskip 3pt
\ligne{\hbox to 40pt{weights\hfill} 
\hbox to \epaiss{\hfill 0\hfill} \hbox to \epaiss{\hfill 1\hfill} 
\hbox to \epaiss{\hfill\ $\,$ 4\hfill}
\hbox to \epaiss{\hfill\ \ \ 12\hfill} \hbox to \epaiss{\hfill\ \ \ \ 29\hfill} 
\hbox to \epaiss{\hfill \ \ \ 34\hfill} %box to \epaiss{\hfill\ \ \ \ 149\hfill}
\hfill
}
}
\hfill}
\vskip 5pt
   The construction of the table required several constraints. The obvious one is that a 
deterministic cellular automaton requires that a single new state is defined for any couple 
consisting of the state of a cell and of its neighbourhood weight. To satisfy that constraint,
the choice of the weights is not arbitrary. In particular, it was not possible to reduce the
highest weight.

\def\BB{\hbox{\tt B}}
\def\RR{\hbox{\tt R}}
\def\MM{\hbox{\tt M}}
\def\WW{\hbox{\tt W}}

    Note that as far as outside the current state the transition function depends on the 
neighbourhood weight only means that the state does not depend on the position of the states
in the neighbourhood. In particular, it means that automatically, the cellular automaton is
rotation invariant.

   It is the place here to discuss about the choices I did in order to obtain the result stated in
Theorem~\ref{letheo}.

   First, I deal with the number of states. That number is dictated by the necessity to get
two types of locomotive in order to implement the crossings as far as the working of that 
structure requires that the structure is able to discriminate between the two types. In previous 
papers, I did that by introducing the \BB\RR-pattern as a simple locomotive and the 
\BB\RR\RR-pattern as a double locomotive. I tried to do that here too and it occurred that even 
with the weight 500 for the heaviest state, it does  not work. So, I decided to introduce the 
pattern \MM\RR{} as another colour for the locomotive. Front and rear have to be different, also 
different from the blank state \WW, the colour of the immense majority of cells which remain in 
that state thanks to line~0 of the table. Accordingly, that entails at least four states. But it 
is not enough. To signalise the path, we need at least one state:  either used as marking the track
on which the locomotive moves or it is placed around the path in a way to signalise it in a non
ambiguous way. This leads us to five states. One more state is the minimum to introduce the 
distinction required by the different mechanisms used to implement the switches. Indeed, it is not 
possible to rely on the neighbourhood: a cell of the track has at least two neighbours belonging 
to the tracks So that sometimes, four to five cells remain free around a cell of the 
track. Moreover, the track crosses the structures needed to implement the switches.

   That discussion leads us to the weights. Their values were chosen in order to, as much as 
possible, distinguish between the states. I postpone the discussion to Section~\ref{stable} as far
as the choice of the weights was motivated after scrutinising Tables~\ref{t_tracks} up 
to~\ref{t_prog}.
%As a few of them can be present at most five times 
%%around a cell, I defined the weights with a clear distance between them. Later, I tried to reduce 
%that distance, especially with the heaviest one. The heaviest one works with 500 as its value. I 
%succeeded to lower it down to 350 but further, 325 for example. After some work detailed in 
%\cite{mmarXiv23b}, I succeeded to reduce the highest weight to 261. It does not work for the 
%immediate lower values.

\section{Main lines of the computation}\label{scenario}

    In the present paper, as we go back to weak universality, we take the general
frame of previous papers of the author, see~\cite{mmarXiv23a} for references to those papers.
Also, we refer the reader to~\cite{mmarXiv23b,mmarXiv23c} for detailed explanations of the 
implementation.

\subsection{The railway model}\label{railway}

    The simulation is based on the railway model devised in~\cite{stewart} which lives 
in the Euclidean plane. It consists of {\bf tracks} and {\bf switches} and the 
configuration of all switches at time~$t$ defines the configuration of the computation 
at that time. There are three kinds of switches, illustrated by Figure~\ref{switches}. 
The changes of the switch configurations are performed by a locomotive which runs over 
the circuit defined by the tracks and their connections organised by the switches.

A switch gathers three tracks $a$, $b$ and~$c$ at a point. In an active crossing,
the locomotive goes from~$a$ either to~$b$ or to~$c$. In a passive crossing, it goes to~$a$
either from~$b$ or from~$c$.
\vskip 10pt
\vtop{
\ligne{\hfill
\includegraphics[scale=0.8]{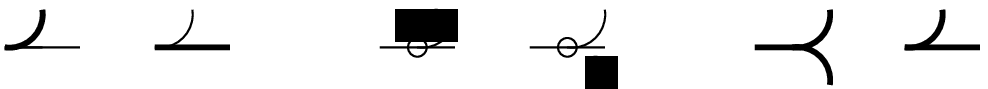}
\hfill}
%\vspace{-20pt}
\begin{fig}\label{switches}
\leurre
The switches used in the railway circuit of the model. To left, the fixed switch, in the
middle, the flip-flop switch, to right the memory switch. In the flip-flop switch, the 
bullet indicates which track has to be taken.
\end{fig}
}

In the fixed switch, the locomotive goes from~$a$ to always the same track: either~$b$ or~$c$
which is called the {\bf selected track}. The passive crossing of the fixed switch is possible and
does not change the selected track. The flip-flop switch is always crossed actively 
only. If the locomotive is sent from~$a$ to~$b$, to~$c$ by the switch, it will be sent to~$c$, 
to~$b$ respectively at the next passage. The memory switch can be crossed actively or passively. 
However, the track taken by the locomotive in an active passage is the track taken by the locomotive
in the last passive crossing. At initial configuration, the crossing of the memory switches is
fixed by the configuration.

Figure~\ref{basicelem} illustrates the circuit which stores a one-bit information. The locomotive 
may enter the circuit either through the gate~$R$ or through the gate~$W$.

  If it enters through the gate~$R$ where a memory switch sits, it goes either through
the track marked with~1 or through the track marked with~0. When it crossed the switch
through track~1, 0, it leaves the unit through the gate~$B_1$, $B_0$ respectively.
Note that on both ways, there are fixed switch sending the locomotive to the appropriate
gate~$B_i$. Note that when the locomotive leaves the unit, no switch was changed.
If the locomotive enters the unit through the gate~$W$, it is sent to the 
gate~$R$, either through track~0 or track~1 from~$W$. Accordingly, the locomotive
arrives to~$R$ where it crosses the switch passively, leaving the unit through the 
gate~$E$ thanks to a fixed switch leading to that latter gate. When the locomotive 
took track~0, 1 from~$W$, the switch after that indicates track~1, 0 respectively and the 
locomotive arrives at~$R$ through track~1, 0 of~$R$. The tracks are numbered according to 
the value stored in the unit. Note that when the locomotive leaves the unit, two switches
were changed: the flip-flop at~$W$ and the memory switch at~$R$. 

\vtop{
\ligne{\hfill
\includegraphics[scale=0.6]{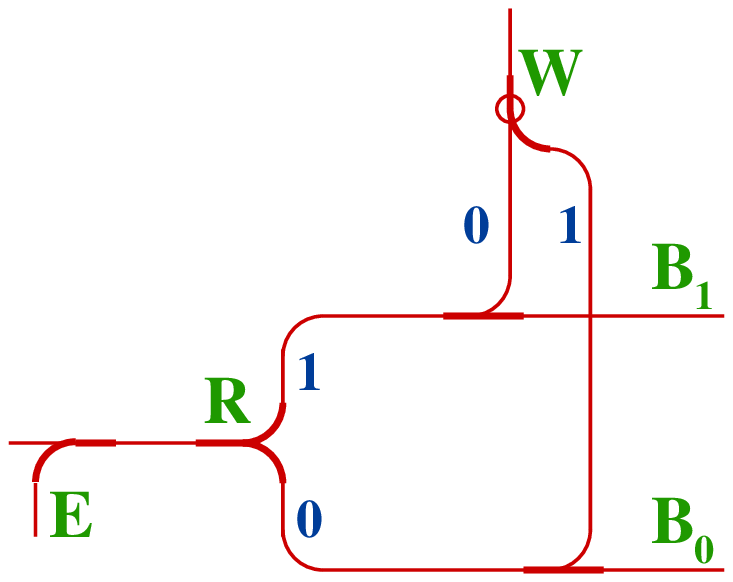}
\hfill}
%\vspace{-20pt}
\begin{fig}\label{basicelem}
\leurre
The basic element containing one bit of information.
\end{fig}
}

By definition, the unit is~0, 1 when both tracks from~$W$ and from~$R$ are~0, 1 
respectively. So that, as seen from that study, the entry through~$R$ performs a reading 
of the unit while the entry through~$W$, changes the unit from~0 to~1 or from~1 to~0: the 
entry through~$W$ should be used when it is needed to
change the content of the unit and only in that case. The structure works like a memory
which can be read or rewritten. It is the reason why we call it the {\bf one-bit memory}.

   We shall see how to combine one-bit memories in the next sub-section as far as we 
introduce several changes to the original setting for the reasons we indicate there.

\subsection{Tuning the railway model}\label{newrailway}

   We first look at the implementation of the tracks in Sub-subsection~\ref{ssstracks}
and how it is possible to define the crossing of two tracks.
In Sub-subsection~\ref{sssauxil} we consider preliminary structures from which we define the
switches described in Sub-subsection~\ref{sssswitch}.
Then, in Sub-subsection~\ref{sssunit}, we see how the one-bit memory is implemented in 
the new context and then, in Sub-section~\ref{sssregdisp}, how we use it in various 
places. At last but not the least, we shall indicate how registers are implemented
in Sub-subsection~\ref{sssreg}.

\subsubsection{The tracks}\label{ssstracks}

    The tracks play a key role in the computation, as important as instructions and 
registers: indeed, they convey information without which any computation is impossible.
Moreover, as can be seen in many papers of the author, that one included, it is not an
obvious issue which must always be addressed.
  
    It is not useful to list the similarities and the distinctions between the present 
implementation and those of my previous papers. The best is to focus on the implementation
used by the present paper. If the reader is interested by the comparison with previous 
implementations the references already indicated give him/her access to the corresponding
papers.

\def\ftt #1 {{\footnotesize\tt#1}}

\def\GG{\hbox{\tt G}}
\def\YY{\hbox{\tt Y}}
\def\VV{\hbox{\tt V}}
   In the present paper and in the case of the heptagrid we shall consider that the tracks are one 
way. As far as presently the locomotive consists of two consecutive cells, the front one and the 
rear, one way is not mandatory but it will be more convenient. By construction, the rear is 
{\bf red}, state {\tt R}. The front may be either {\bf blue}, state \BB, or {\bf mauve}, 
state \MM{}. The reason of two kinds of locomotive will be explained later.
  
  Here too, the elements of tracks consist a \YY-cell, most often, having two \YY-neighbours
from side to side of the cell. Note that the weight of \YY{} is~1.
%two non-blank neighbours, \GG{} and \YY{} whose respective weights are 48 and 144.  
The structure is illustrated by Figure~\ref{fel_tr_i}.
\vskip 5pt
\vtop{
\ligne{\hfill\hskip-20pt
\includegraphics[scale=1.8]{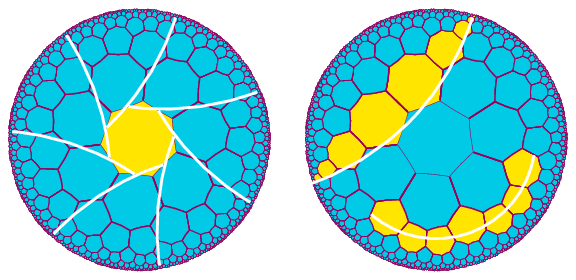}
\hfill}
\begin{fig}\label{fel_tr_i}
\leurre
To left, a single element a track. To right, two examples of tracks: one of them following a line,
the other, an arc of a circle. To left, note the rays delimiting the sectors in 
order to facilitate the location of the cells.
\end{fig}
}
\vskip 5pt
The left-hand side 
picture of Figure~\ref{fel_tr_i} illustrates an element of a track: a \YY-cell.
The left-hand side part of the figure shows us the sectors we use to explain the right-hand side
picture. In that part of the figure we show two types of tracks: one of them follows a line of the 
hyperbolic plane, here a mid-point line and the other follows an arc of a circle of the
hyperbolic plane. The supports of the cell constituting the tracks along the line are, from left
to right: (3,54), (3,20), (3,7), (3,2), (2,1), (1,1),(1,2), (1,5), (1,13) and (1,34). The other
cells of the track are not visible in the figure. For the arc of a circle, the tiles are, again from
left to right: (4,2), (4,3), (4,4), (5,2), (5,3), (5,4) and (6,2). Later, Figure~\ref{f_rac_i}
presents two segments of lines joined by an arc of a circle.

\def\FF{{\tt F}}
The motion is organised according to the following scheme:
\vskip 5pt
\ligne{\hfill{\tt WWWW\hskip 20pt FWWW\hskip 20pt RFWW\hskip 20pt WRFW\hskip 20pt WWRF\hskip 20pt
WWWR}
\hfill(\numerrel)\hskip 15pt}
\vskip 5pt
\noindent
were \FF{} denotes the front cell and \RR{} denotes the rear one. The motion depends on the fact 
that \FF{} and \RR{} receive different weights, which allows the motion to take place. As far as 
the total weight of \WW{} and \FF{} is involved and that weight only, the motion may  occur in 
both directions on the track. However, we require that a single locomotive moves on a given track 
at a considered moment and that all tracks are one-way. As far as both versions of \FF{} have a 
weight which is different from that of \RR, both kinds of locomotive can move. The tracks will be 
organised along arcs of circles and along segments of lines.

\vskip 5pt
\vtop{
\ligne{\hfill
\includegraphics[scale=1.5]{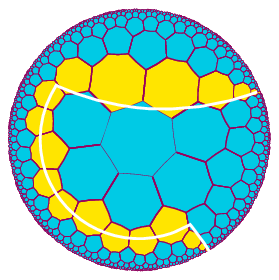}
\hfill}
\begin{fig}\label{f_rac_i}
\leurre
The idle configuration of a path where an arc of a circle joins tow segments of line. 
Here, the radius of the circle supporting the arc of the \YY-tiles is $2$. 
\end{fig}
}

\vskip -5pt
\vtop{
\ligne{\hfill
\includegraphics[scale=0.55]{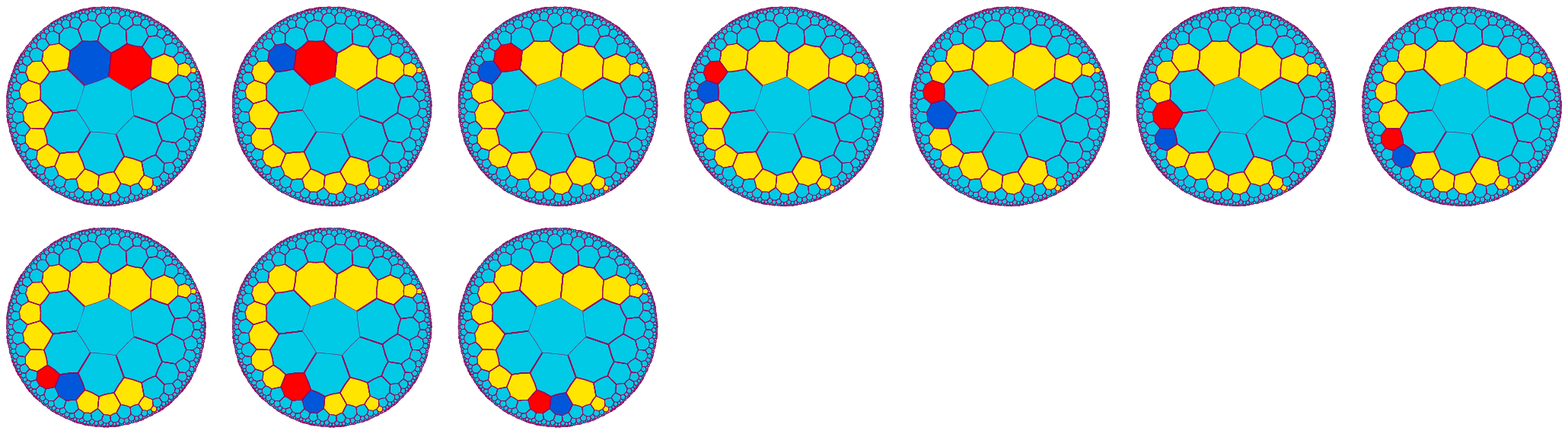}
\hfill}
\vskip-10pt
\ligne{\hfill
\includegraphics[scale=0.55]{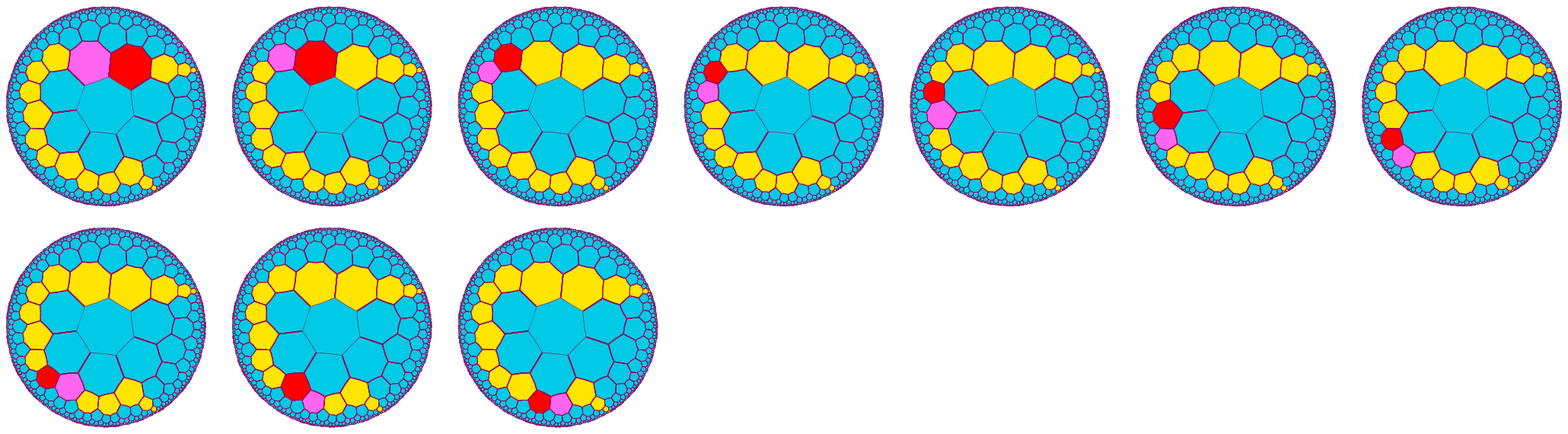}
\hfill}
\vspace{-10pt}
\begin{fig}\label{f_rac_m}
\leurre
Top two rows: motion of a blue locomotive on the track of Figure~{\rm\ref{f_rac_i}}.
Bottom rows: motion of a mauve locomotive on the same track.
\end{fig}
}
\vskip 5pt

Figure~\ref{f_rac_i} illustrates a path where an arc of a circle joins two segments of line. The 
arc of a circle goes from (2,2) to (5,2), the \YY-cell being on a circle of radius~~2 around the
central cell. One segment goes through (2,2), (1,1), (7,1), (7,2), (7,5), (7,13) and (7,34).
The other segment of lines goes through (5,2), (5,7), (5,20) and (5,54).

\vskip 5pt
Figure~\ref{f_rac_m} illustrates the motion of a blue locomotive, top two rows, and that of a mauve
one, bottom two rows, on the track just defined and illustrated by Figure~\ref{f_rac_i}. 
Figure~\ref{f_rac_i} illustrates the notion of {\bf idle configuration}: it is
a window in which the locomotive is not present in a disc of radius~3 around cell~0.

\subsubsection{Auxiliary structures and crossings}\label{sssauxil}

   In order to implement switches in our setting, we need auxiliary structures we already introduced
in some other papers. Those structures are the fork, the converters and the filters whose idle 
configurations are given by Figure~\ref{f_auxil}. 

\vskip 5pt
\vtop{
\ligne{\hfill
\includegraphics[scale=0.95]{secteurs.ps}
\includegraphics[scale=0.95]{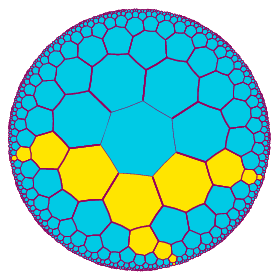}
\includegraphics[scale=0.95]{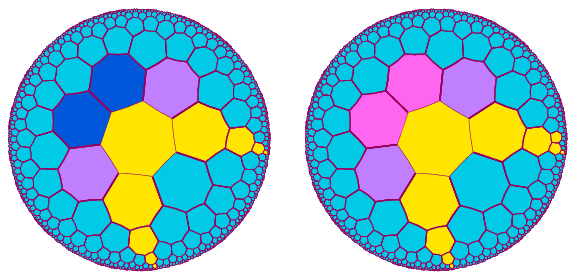}
\hfill}
\ligne{\hfill
\includegraphics[scale=0.95]{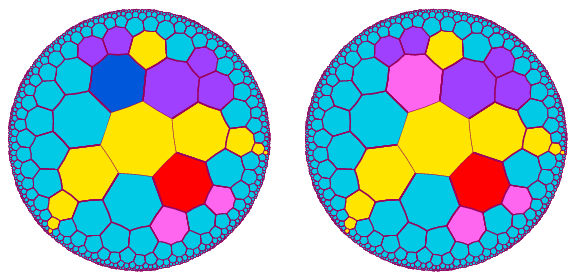}
\hfill}
\vskip-20pt
\begin{fig}\label{f_auxil}
\leurre
Idle configurations: top row: from left to right, the sectors, the fork and two converters.
Bottom row, filters. The sectors allow us to locate the cells in the configurations. The track 
consists of the \YY-cells.
\end{fig}
}

The fork allows the duplication of a locomotive, whichever its colour. The converters change a 
blue locomotive to a mauve one and conversely: the converter with two \BB-cells changes a mauve 
locomotive to a blue one and that with two \MM-cells performs the opposite conversion. The figures 
are focused on a disc of radius~3 around a central cell~$c$: it is a {\bf window} which allows us 
to see what happens in the neighbourhood of~$c$.

The cells of the tracks joining at the fork are: (5,40), (5,15), (5,5) and (5,2) for the 
arriving track at (4,1), then (3,1), (3,2), (3,6), (3,17) and (3,46) for the left-hand side track 
leaving the fork, and then (5,1), (6,2), (6,6), (6,17) and (6,46) for the right-hand side track 
leaving that structure, see the leftmost picture in the top row of Figure~\ref{f_auxil} to locate 
the cells. In the figures, the cells of the track consists of \YY-cells.

Figure~\ref{f_frk_m} illustrates the move of a locomotive through the fork: top row, a blue one, 
bottom row, a mauve one. Note how two locomotives are created from the arriving one in the 
neighbourhood of cell~0.
\vskip 5pt
\vtop{
\ligne{\hfill
\includegraphics[scale=0.6]{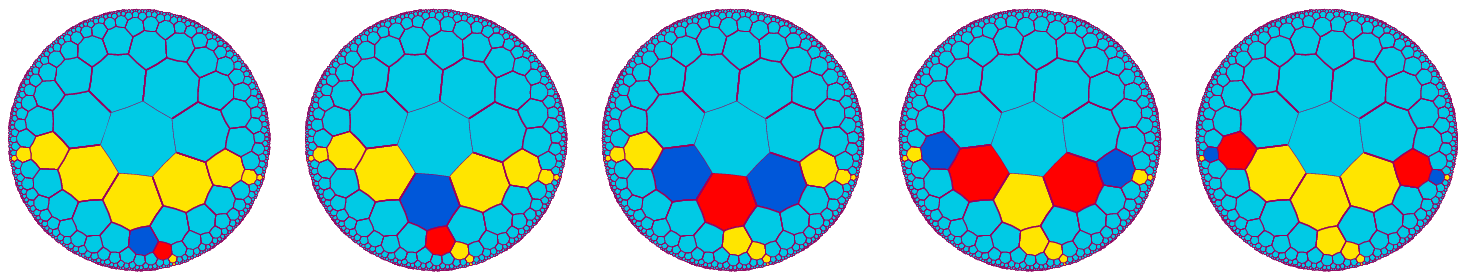}
\hfill}
\ligne{\hfill
\includegraphics[scale=0.6]{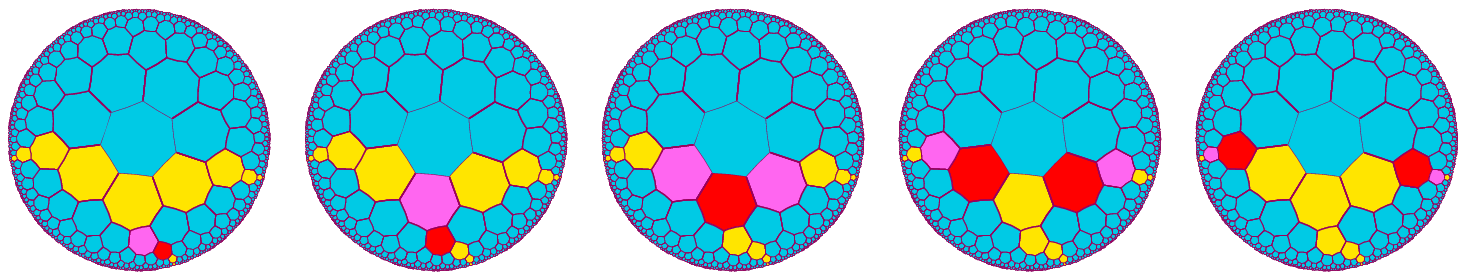}
\hfill}
\vskip-20pt
\begin{fig}\label{f_frk_m}
\leurre
Motion of a locomotive through the fork. Top: a blue one; bottom: a mauve one.
\end{fig}
} 
\vskip 10pt
The converter changes the colour of a locomotive into the other colour. The track is defined by 
the cells (6,21), (6,8), (6,3), (6,1), 0, (4,1), (4,4), (4,12) and (4,33). The converter consists
of two \VV-cells and two cells of the colour to which the converter changes the arriving locomotive
whose colour is assumed to be the opposite colour to that of the converter. The \VV-cells are at
the tiles (3,1) and (7,1) while the tiles which are both either \BB- or \MM- are (1,1) and (2,1).
Figure~\ref{f_ch_m} shows us that the change of colour for the front of the locomotive is performed
at cell~0 as far as that cell can see the front of the locomotive when it arrives at the cell (6,1).

\vtop{
\ligne{\hfill
\includegraphics[scale=0.6]{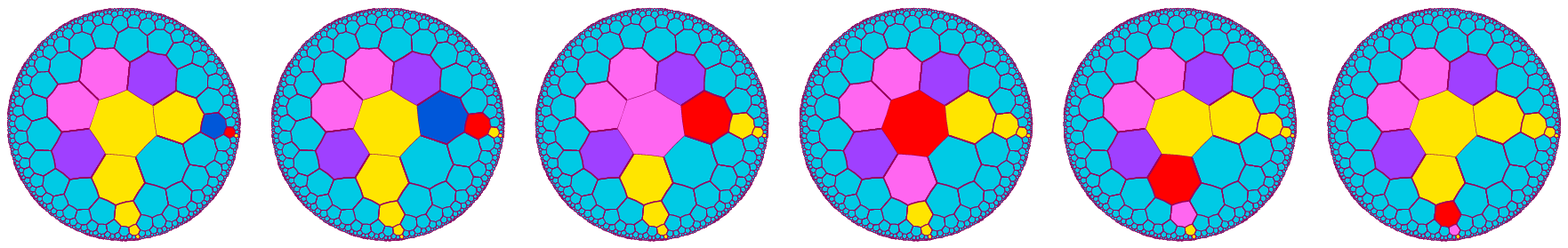}
\hfill}
\ligne{\hfill
\includegraphics[scale=0.6]{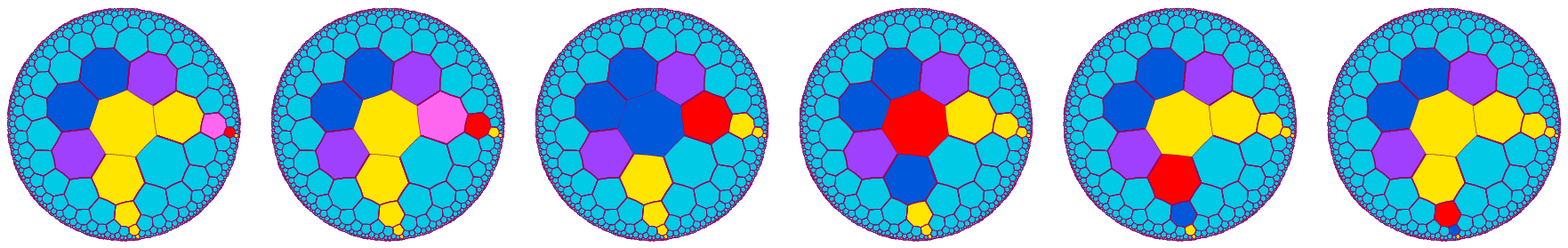}
\hfill}
\vskip-20pt
\begin{fig}\label{f_ch_m}
\leurre
Top row: the structure changes a blue locomotive into a mauve one. 
Bottom row : conversely, the change of a mauve locomotive into a blue one.
\end{fig}
} 

The filters allow the passage of the locomotive of a given colour and prevents the passage for
the locomotive of the other colour. The authorised colour is represented by that of the cell (1,1).
The track follows a segment of a line, from right to left: (6,21), (6,8), (6,3), (6,1), (1,1), 
(3,1), (3,4), (3,12) and (3,33). The central cell has three non \WW- and non \YY-neighbours: a 
\VV-one at~(7,1), an \RR-one at (5,1) and the colour of the filter, either \BB- or \MM-n at (1,1). 
Moreover, for technical reasons, cells (1,1) and (7,1) are decorated by \VV-cells at (1,3), (1,4) 
and at (7,2), (7,3) respectively. The \RR-cell at (5,1) is decorated by two \MM-cells at (5,2) and 
(5,4). At last and not the least a \YY-cell occurs at (1,2), a neighbour of the cell (1,1).

Figure~\ref{f_sel_m} illustrates the move through a filter.
%whose colour is determined by the cell (1,1) which is either \BB- or \MM. 
The filter let the locomotive go on its way if it is of the same colour and it stops it when the 
colour is different. Note the configurations of cell~0 when the filter stops a locomotive. The 
stopping is obtained as far as cell~0 always remains \YY- when the locomotive has not the required 
colour. The front vanishes and then the rear does the same.
\vskip 5pt
\vtop{
\ligne{\hfill
\includegraphics[scale=0.6]{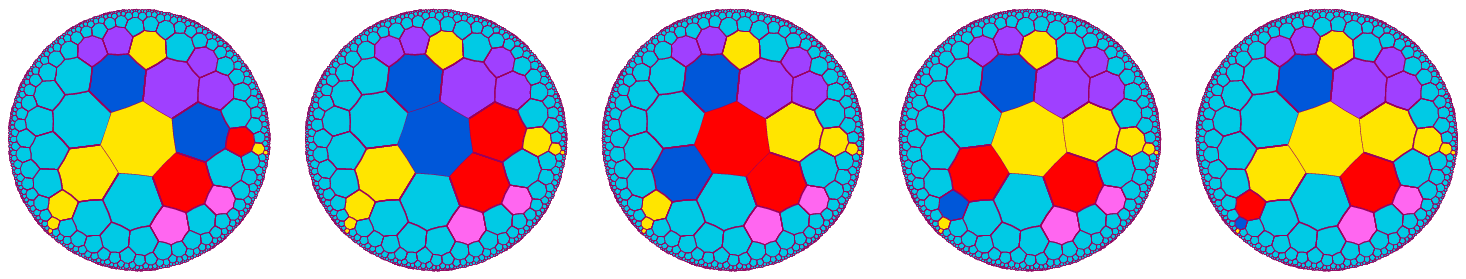}
\hfill}
\ligne{\hfill
\includegraphics[scale=0.6]{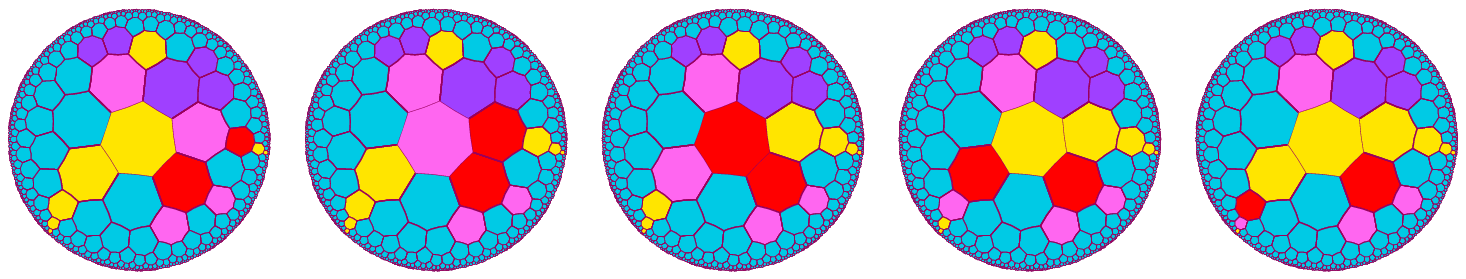}
\hfill}
\ligne{\hfill
\includegraphics[scale=0.6]{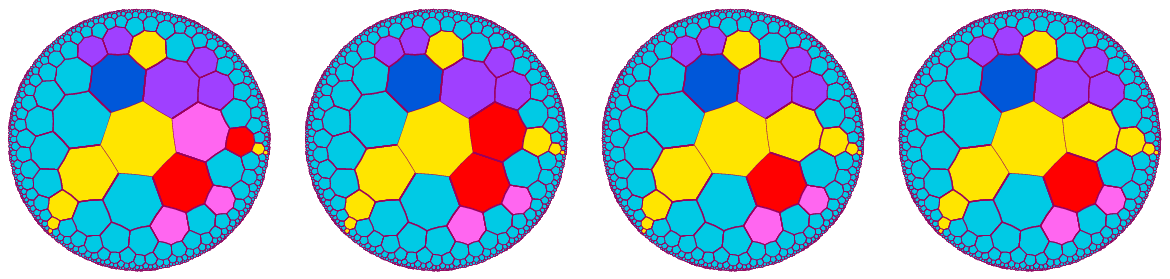}
\hfill}
\ligne{\hfill
\includegraphics[scale=0.6]{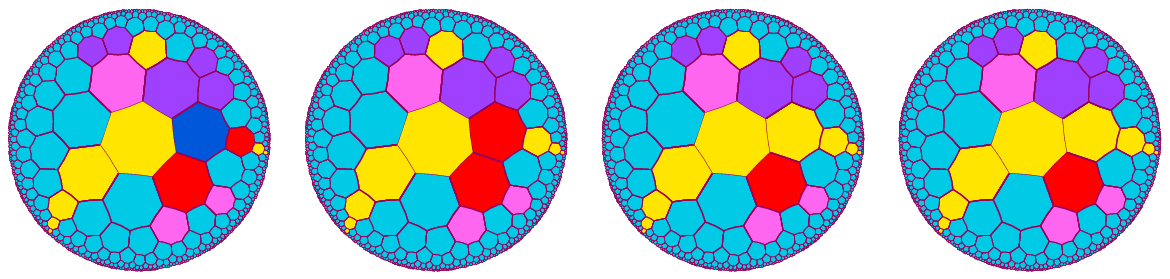}
\hfill}
\vskip-20pt
\begin{fig}\label{f_sel_m}
\leurre
Top rows: the filter let the locomotive go on its motion, topmost, the blue locomotive,
below, the mauve locomotive.
Bottom rows: the filter stops the locomotive of the other colour.
\end{fig}
}

In section~\ref{stable} we give the entries of the table ruling those different motions. Many 
entries are used for that purpose.

\subsubsection{The switches}\label{sssswitch}

The section follows the implementation described in~\cite{mmbook2} for instance. We 
reproduce it here for the reader's convenience. However, it is much simplified in the case of 
several structures. The illustrations of the section show us a window when it is possible to do so 
or a diagram which illustrates how auxiliary structures are combined to constitute a switch. When a 
window can be used, we give the picture illustrating the sectors and the idle configuration. Another
figure illustrates the motion of a locomotive across the window.

We start the study of the switches by the fixed switch. Remember that such a switch can be crossed
actively or passively. As the tracks we consider are all one-way tracks, there is no need of a
switch in an active crossing of a fixed switch. When it is needed, the correspondence of the active
switch with the passive will be explained. Presently, we consider the passive switch only whose
idle configuration is illustrated by Figure~\ref{f_fix_i}.

\def\hww{\hskip-5pt}
\def\hw{\hskip-4pt}

\vskip 5pt

\vtop{
\ligne{\hfill
\includegraphics[scale=2]{secteurs.ps}
\includegraphics[scale=2]{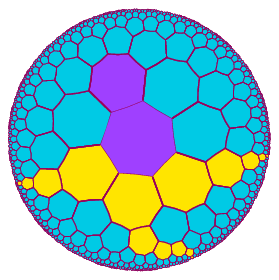}
\hfill}
\vskip-20pt
\begin{fig}\label{f_fix_i}
\leurre
To left, the sectors, to right, the idle configuration of a fixed passive switch.
\end{fig}
}

\vtop{
\ligne{\hfill
\includegraphics[scale=0.6]{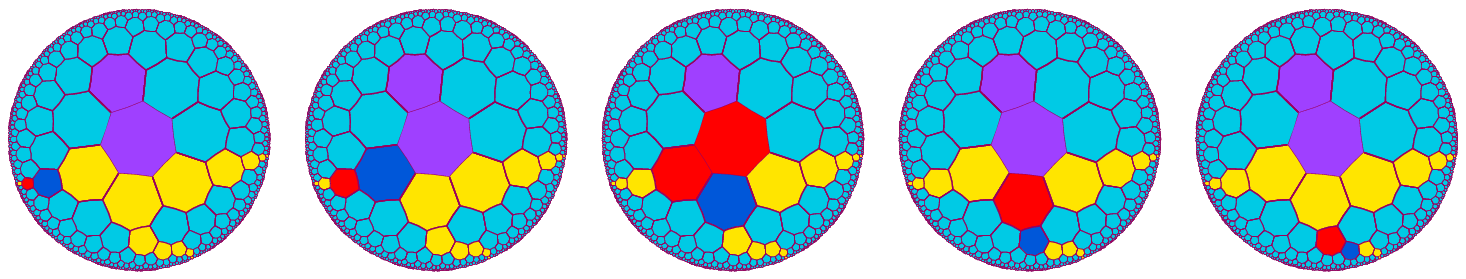}
\hfill}
\ligne{\hfill
\includegraphics[scale=0.6]{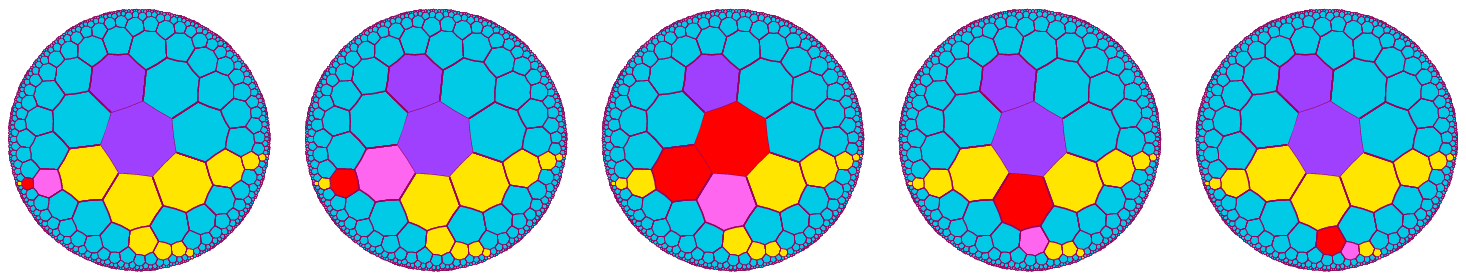}
\hfill}
\ligne{\hfill
\includegraphics[scale=0.6]{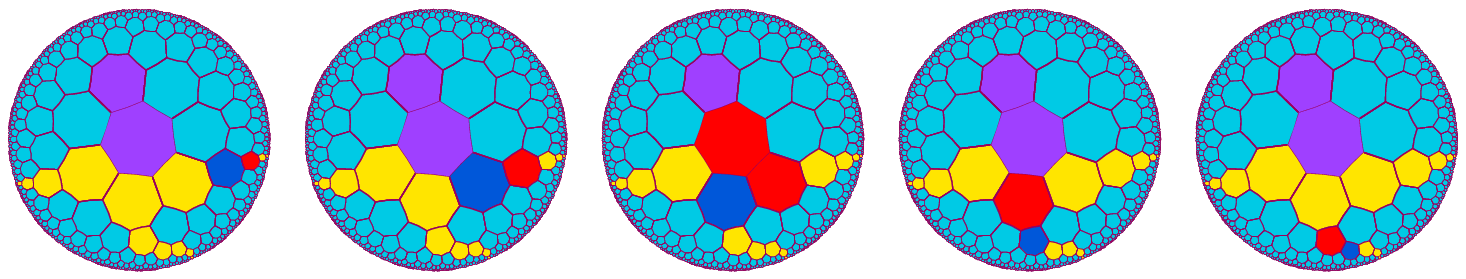}
\hfill}
\ligne{\hfill
\includegraphics[scale=0.6]{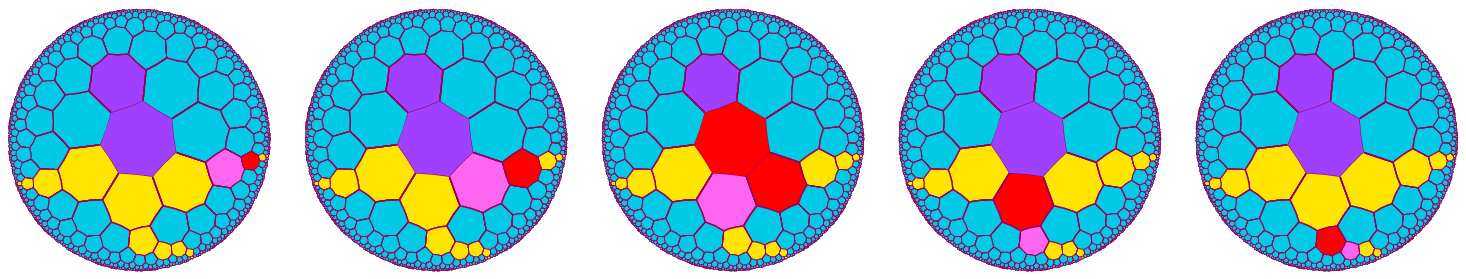}
\hfill}
\vskip-20pt
\begin{fig}\label{f_fix_m}
\leurre
The passive crossing of the fixed switch. Top rows, the crossing through the left-hand side branch,
bottom rows, the crossing through the right-hand side branch.
\end{fig}
}

%\vskip 10pt
   The left-hand side part of Figure~\ref{f_fix_i} allows the reader to locate the cells of the 
right-hand side part of the figure, here, the idle configuration of the fixed switch. 
   Note that the two tracks arriving at cell~0 are defined by the following cells:
(3,55), (3,21), (3,8), (3,3) and (3,1) from the left-hand side together with the
cells (6,54), (6,20), (6,7), (6,2) and (5,1) from the right-hand side. 
The exit track is defined by the cells (4,1), (4,4), (5,5), (5,6), (5,18) and (5,48).
That can be checked on the right-hand side part of Figure~\ref{f_fix_i} where the cells of the 
track are \YY-cells. Note the difference of configuration with the fork. 

In the fork, there is no special cell at tile~0 which is simply a \YY-cell. In the fixed switch,
the central cell is a \VV-cell. It is decorated with another one in order to distinguish that
cell from other \VV-cells which occur in other configurations. The role of the central \VV-cell
is to allow the locomotive arriving at (4,1) from (3,1) to go to (4,4) and the further \YY-cells
and, at the same time, to prevent the locomotive to go on to (5,1) and the further \YY-cells.
It is what would happen if the central cell would be \YY: we would get a fork.
Preventing the locomotive from going to (5,1) is obtained by the change of colour of 
the cell (1,1): when a locomotive arrives to (3,1) or (5,1), the \VV-cell at~0 becomes an \RR-cell 
for one step of the computation which is enough for our purpose. From the definition of a weighted 
cellular automaton, it is plain that if the table works for locomotives coming from the left-hand 
side branch it also works for locomotives coming from the right-hand side branch.

With the structure we have gathered up to now, we can describe how crossings are implemented.
With a rather large number of states, it is possible to implement direct crossings, see for
instance~\cite{fhmmTCS}. When the number of states is relatively small as the case is in previous
papers it is needed to associate auxiliary structures in a complex way. Here, it is possible to
implement an almost direct crossing as illustrated by Figure~\ref{f_cross}.

\vskip 5pt
\ligne{\hfill
\vtop{\leftskip 0pt\parindent 0pt\hsize=250pt
\ligne{\hfill
\includegraphics[scale=0.45]{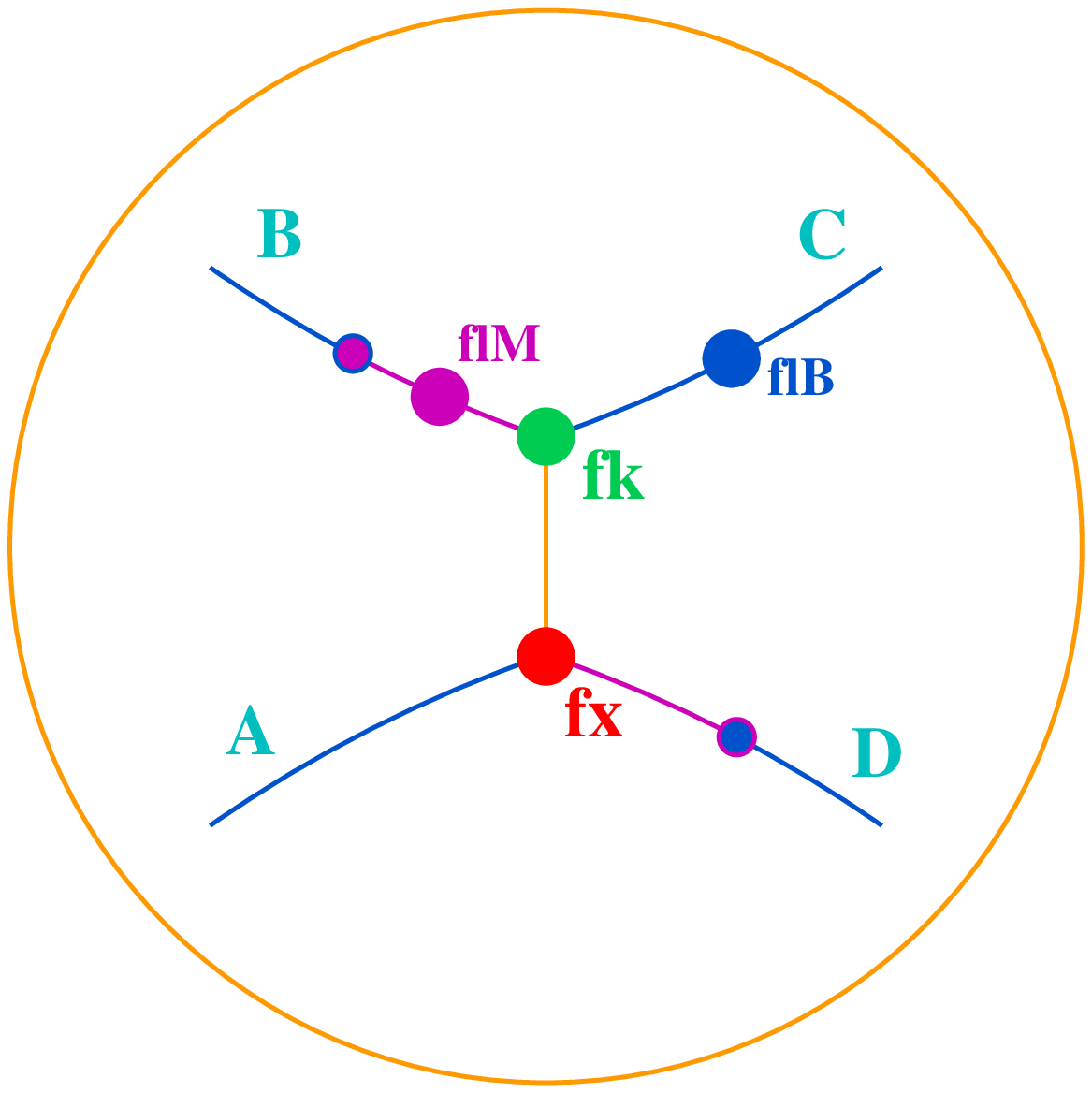}
\hfill}
\vspace{-20pt}
\begin{fig}\label{f_cross}
\leurre
The structure of a crossing.
\end{fig}
}
\hfill}
\vskip 5pt
The idea is to use both the existence of two types of locomotives and the filters which allow
a locomotive of a definite colour to go on its way and only it. The required working is facilitated 
by the fact that the difference of two types of locomotive occurs on the front of it. The idea is 
to associate one track with the blue locomotives and the other with the mauve ones. By associating 
a fixed switch with a fork we obtain the configuration of Figure~\ref{f_cross}. In the crossing of 
the figure, a locomotive arriving from~$A$ is supposed to go onto the $C$-branch while a locomotive 
arriving from~$D$ is supposed to go to the $B$-one. On the track coming from~$D$ and before the 
fixed switch, we put a converter which changes a blue locomotive into a mauve one. 

Assume for a while that one track is blue and the other is mauve. By those words we mean that
on a blue, mauve track a \BB-, an \MM-locomotive is assumed to run and not an otherwise coloured 
one. After the fork, it is enough to place a \BB-filter on the blue track and and \MM-one on the 
mauve track. Accordingly the locomotive of the appropriate colour is allowed to go further. We may 
extend that situation to any crossing: either the two tracks are of the same colour or of opposite 
colour. When the tracks have the same colour, it is enough to change the colour of the locomotive on
one track by giving it the opposite colour and to restore the required colour after the crossing of
the fork. As far as it is possible to change any locomotive in a locomotive of the opposite 
colour, we may perform any crossing according to what we already said. Figure~\ref{f_cross} 
illustrates the case of a crossing of two blue tracks.

For the crossing, we can use the filters as fixed structure. However to implement the other switches
we need to be able to change the working of the filter. We need programmable structures. In
fact, the filter as illustrated in Figure~\ref{f_auxil} can be used to that purpose.  
Figure~\ref{f_flt_m} illustrates how we can perform the change of a filter.

A close look at Figure~\ref{f_auxil} shows us one element of track just above the cell 
indicating the colour chosen by the filter. Accordingly, a locomotive may arrive at that point
provided a track arrives there. We assume that a mauve locomotive arrives and it creates a \WW-cell
at (1,2) which is in contact with (1,1) where the cell bears the colour accepted by the filter.
That \WW-cells reduces the neighbouring weight of (1,1) so that it changes its colour: a \BB-one
becomes \MM- while an \MM-one becomes \BB-. The motion is illustrated by Figure~\ref{f_flt_m}.

\vskip 5pt
\ligne{\hfill
\vtop{\leftskip 0pt\parindent 20pt\hsize=330pt
\ligne{\hfill
\includegraphics[scale=0.5]{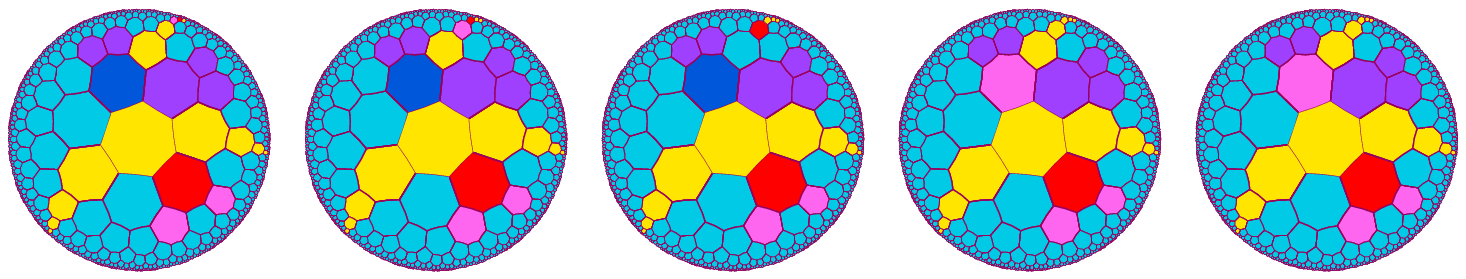}
\hfill}
\ligne{\hfill
\includegraphics[scale=0.5]{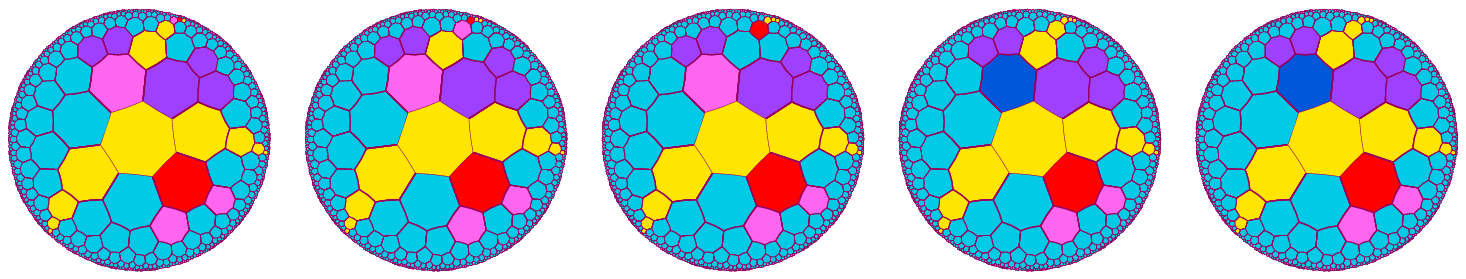}
\hfill}
%\vspace{-10pt}
%\ligne{\hskip 209pt \vrule height 12pt depth 2pt width 2pt\hfill}
\begin{fig}\label{f_flt_m}
\leurre
Top, the mauve locomotive changes a blue filter into a mauve one. Bottom, a mauve locomotive again
changes the mauve filter into a blue one.
\end{fig}
}
\hfill}
\vskip 5pt
We are now ready to investigate the flip-flop and the memory switches.

The flip-flop switch and both parts of the memory switch require a much more involved
situation. The global view of an idle configuration of the flip-flop is illustrated by 
Figure~\ref{f_basc}. The switch is crossed actively only and we also impose it is crossed by a blue
locomotive only. A fork operates the first action of the switch: the arriving locomotive is
duplicated as a copy on each branch leaving the fork. On one branch a blue filter sits: it lets the
locomotive go on along the track: it is the selected branch. On the other branch, the locomotive is
stopped by a mauve filter.

\vskip 10pt
\vtop{
\ligne{\hfill
\includegraphics[scale=0.45]{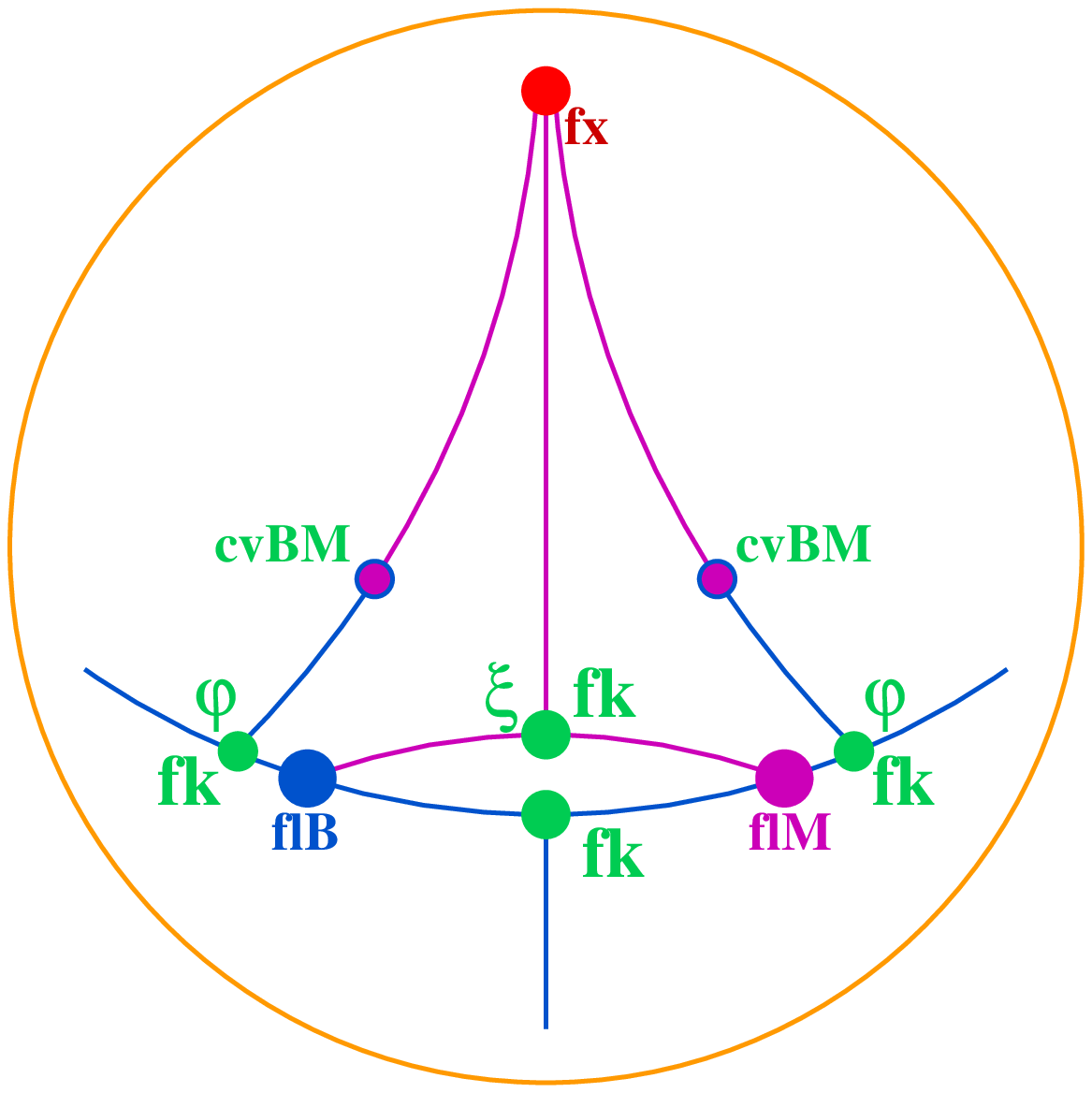}
\hfill}
\vspace{-30pt}
\begin{fig}\label{f_basc}
\leurre
Scheme of the implementation of a flip-flop switch. Note the filters. Note the mauve tracks:
they are segment of lines, not arc of circles.
\end{fig}
}

However, once the locomotive crossed the switch, the selected branch must change. That action is 
performed as follows. Before the blue filter, there is a fork~$\varphi$. One branch of it is the 
track defined by the switch. The other branch of the fork leads to a converter which converts the 
blue locomotive running through it into a mauve one. After the converter, the track on which the 
mauve locomotive runs arrives at a fork $\xi$ whose branches $\beta_\ell$ and $\beta_r$ reach the 
filters. As far as the filters are reached by a mauve locomotive on the appropriate track, the 
filters exchange their colours: the blue filter becomes mauve and the mauve one becomes blue. 
Accordingly, the selected track is changed. Of course, the mauve filter is followed by a fork 
which operates symmetrically with respect to~$\varphi$: on the other branch, there is a converter
from blue to mauve and the mauve track arrives to~$\xi$ too thanks to a fixed switch as 
illustrated by Figure~\ref{f_basc}.

   Presently, we turn to the memory switch, active and passive parts.

   We first deal with the active part. It looks like the flip-flop switch with this 
Figure~\ref{f_memo} illustrates both parts of the memory switch: to left, the active part of the 
switch, to right, its passive part.

The active switch looks like a simplified flip-flop switch. The action on the filters is
operated from a fork~$S$ on the left-hand side part of Figure~\ref{f_memo}  which may receive a 
mauve locomotive sent from the structure illustrated by the right-hand side part of that figure.
\vskip 10pt
\vtop{
\ligne{\hfill
\includegraphics[scale=0.4]{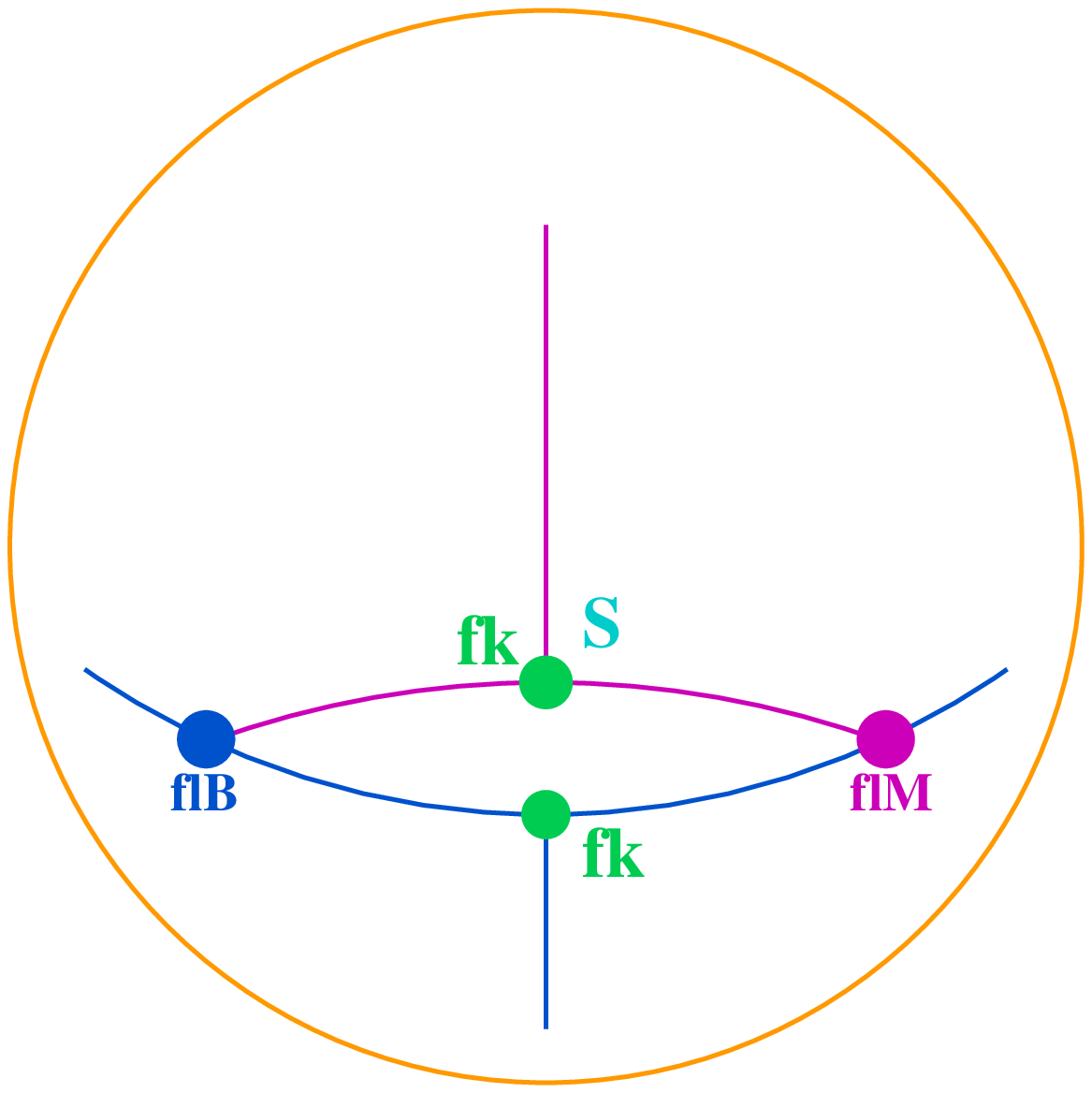}
\includegraphics[scale=0.4]{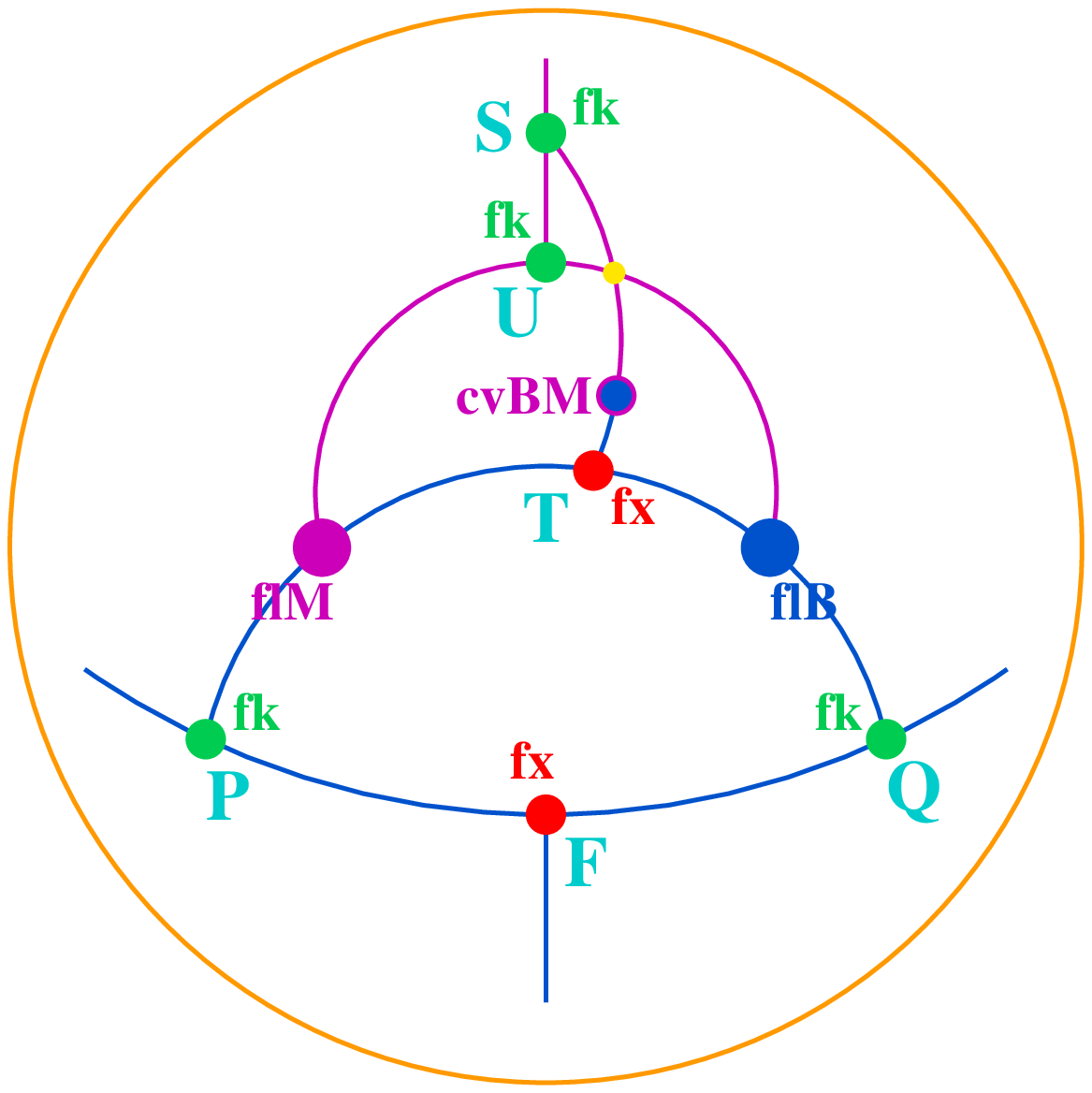}
\hfill}
\vspace{-20pt}
\begin{fig}\label{f_memo}
\leurre
To left, the active memory switch, to right, the passive one.
\end{fig}
}
\vskip 5pt

Let us describe the working of the passive switch. 

The locomotive, a blue one, arrives through~$P$ or through~$Q$, each point sitting from side to 
side of~$F$, as shown on the figure. At $P$ and~$Q$ a fork is sitting. Assume that the locomotive
comes through~$P$, if it comes through~$Q$, the argument is symmetrical. From~$P$, two locomotives 
are sent: one goes to~$F$ and then follows the exit track of the fixed switch sitting at~$F$ while 
the other locomotive is used to change the filters at both switches if it is needed. If the 
selected track is that where $P$ sits, the filter met by the locomotive is mauve, so that it is 
stopped: the configuration of the switches, both its active and passive parts, is unchanged. If 
the side of~$P$ is not the selected branch of the switch, the filter met after $P$ is blue so that 
the locomotive goes on its way. It becomes mauve and it is sent to two forks~$S$, on the active 
switch and $U$ on the passive one and each one of those forks sends a mauve locomotive to the 
filters of both switches in order to change the selected track. Note that on the active switch, 
the selected track bears a blue filter while on the passive switch, the selected track bears a 
mauve one, as illustrated on Figure~\ref{f_memo}. Note the more complex configuration of the 
passive memory switch. We can see a crossing of mauve tracks. It means that mauve locomotive are 
running on those tracks. Now, as already mentioned, such a crossing is dealt with as the crossing 
of tracks where blue locomotives are running. The difference is that on one of the mauve tracks, 
the mauve locomotive becomes blue for a while in order to get a correct crossing. The property of 
using mauve locomotives as signals allows us to use that property of transporting a distinction 
detected in one part of the circuit to another part or to several other parts where the 
distinction need to be reported to.

   Note two other points. First, note that each fixed switch and each fork requires a disc of
radius at least three, and that a circle of radius~3 contains 56 tiles. In fact, the number of tiles
of a circle exponentially rises with the size of its radius. So that the passive memory switch 
requires a huge amount of tiles. The second remark is that in general a single locomotive
operates in the whole circuit of the computation, with the exception of an auxiliary one behaving 
as a signal sent from a part of the circuit to another one. We neglect here the copied locomotive 
which is later destroyed by the appropriate filter. That property means that in between two passages
of the locomotive to the same passive memory switch, there is enough time for the locomotive signal
to operate the change required by the definition of the switch if it is the case to be operated.

\subsubsection{The one-bit memory}\label{sssunit}

It is now time to implement the one-bit memory. Figure~\ref{fonebit} illustrates the
construction.

\def\WWW{{\bf W}}
\def\DD{{\bf D}}
\def\RRR{{\bf R}}
\def\EE{{\bf E}}
\def\ZZ{{\bf Z}}
\def\bbz{{\bf b0}}
\def\bbu{{\bf b1}}
\def\zz{{\bf 0}}
\def\uu{{\bf 1}}

We can see the active memory switch at~\RRR{} and the passive one at~\EE. The dark letters
which stand by the blue circle indicate {\bf gates} of the one-bit memory: \WWW, \RRR, \EE,
\zz{} and \uu. We can easily see that if the locomotive enters the unit through 
the gate~\RRR, where an active memory switch is sitting, then it leaves the memory through the 
gate~\zz{} or through the gate~\uu{} depending on the information stored in the memory: that 
information is provided the unit by the positions of the switch at~\WWW{} and those at~\RRR{} 
and~\EE. Note that the positions at~\RRR{} and at~\EE{} are connected by the path from~\EE{} 
to~\RRR, see the figure.

\vskip 5pt
\ligne{\hfill
\vbox{
\ligne{\hfill
\raise 30pt\hbox{\includegraphics[scale=0.63]{elem_gb.ps}}
\hskip-20pt
\includegraphics[scale=0.45]{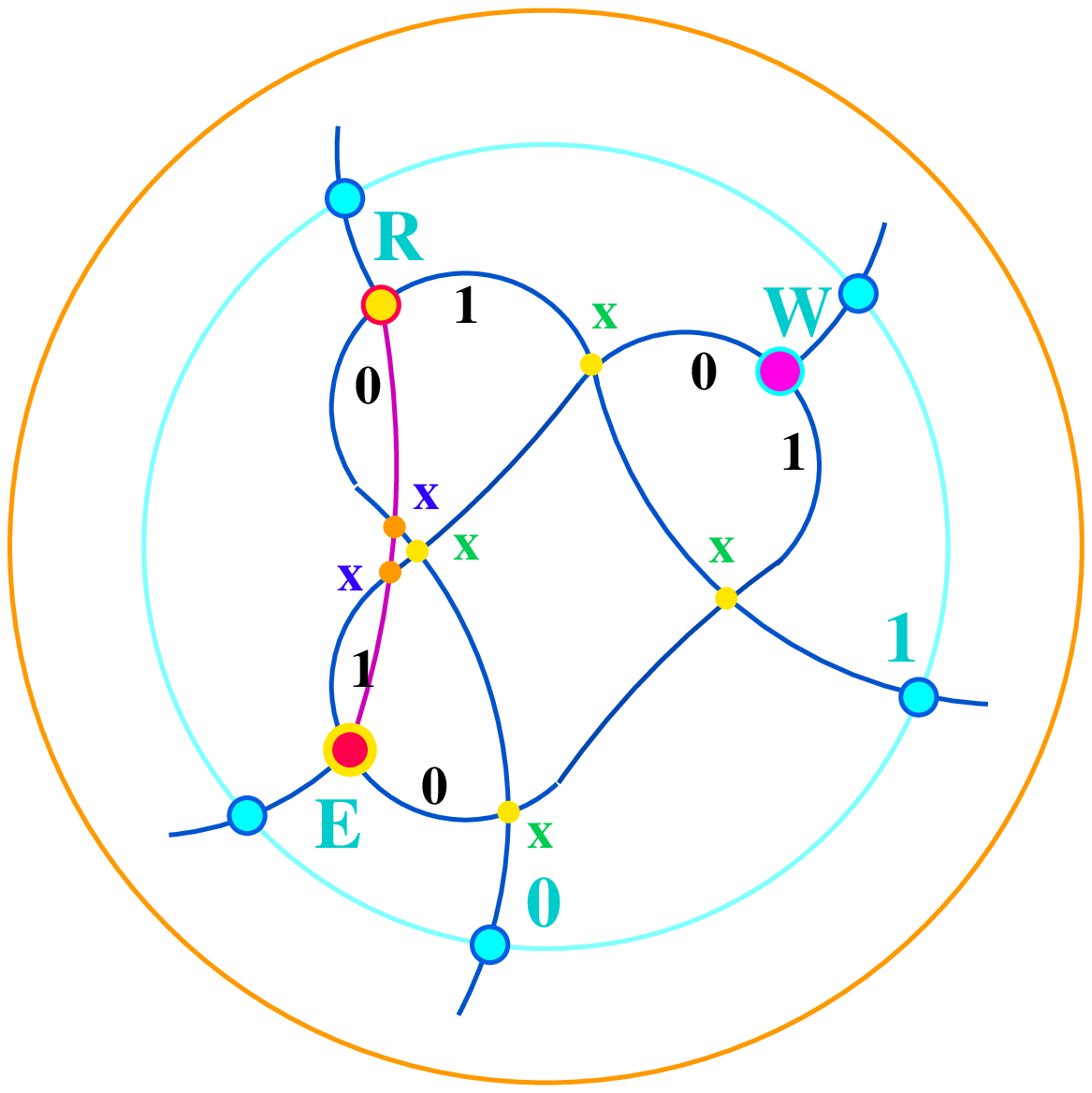}
\hfill}
\vspace{-30pt}
\begin{fig}\label{fonebit}
\leurre
To left,the theoretical structure of a one bit memory: Figure~{\rm\ref{basicelem}} is replicated for
the reader's convenience. To right, the idle configuration of the one-bit memory in the heptagrid. 
Note the four crossings in the implementation. Note that the connection from~\EE{} to~\RRR{} is 
realised by a segment of a line, a shorter path, which is a mauve track as used for a track where a 
mauve locomotive is supposed to be running. Note the crossings of the mauve track with blue ones
which we already discussed.
\end{fig}
}
\hfill}

%\vskip 10pt
When the locomotive enters the memory through the gate~\WWW{} where a flip-flop switch is
sitting, it goes to~\EE{} through one of both tracks leaving the switch. If it goes
through the track marked by~\zz, \uu, it arrives to~\EE, a passive memory switch, by the track 
marked with the opposite symbol, \uu, \zz{} respectively. Indeed, when the locomotive crosses~\WWW,
the passing makes the selected track to be changed so, if it went through one track, after
the passage, in particular when the locomotive arrives at~\EE, the new selected track
at~\WWW{} is the track through which the locomotive did not pass. So that the
track marked by one symbol at~\WWW{} should be marked by the opposite one at~\EE. 
Note that the passive memory switch at~\EE{} and the active memory switch at~\RRR{} correspond to
the memory switch at \RRR{} in Figure~\ref{basicelem} which is reproduced for the convenience of the
reader. It is the reason why those active and passive memory switches are connected by a mauve
track in Figure~\ref{fonebit}. Note that in between two consecutive visits to the same one-bit
memory, it is assumed that there is enough time in order that possible rewriting of a previous
visit are completed.

   As the one-bit memory will be used later, we introduce a simplified notation:
in Figure~\ref{fonebit}, the memory structure is enclosed inside a blue circle. At its 
circumference the gates are labelled by the same symbols as in the hyperbolic picture. In the next 
figures, when a one-bit memory will be used, we shall indicate it by a light blue disc with, at its 
border, the five gates mentioned in Figure~\ref{fonebit}.

\subsubsection{From instructions to registers and back}\label{sssregdisp}

   As will be explained in Sub-subsection~\ref{sssreg}, the locomotive arrives at a 
register at a point which depends on the type of the operation to be performed. 

It depends on the type only, whether it is to decrement or to increment. It does
not depend on namely which instruction of the program required the execution of that
operation. Moreover, the return path of the locomotive once it performed its operation
is the same in most cases. 

Accordingly, when the locomotive goes back from the register
to the program, it is important to define the point at which it will return. 
Correspondingly with what we said, the unique solution is to keep track of that
information before the locomotive enters the register where that information disappears
when the locomotive goes back after performing its instruction.

\def\DDS{${\mathbb D}_S$}
\def\DDI{${\mathbb D}_I$}
\def\DDD{${\mathbb D}_D$}
\def\DDO{${\mathbb D}_O$}
\def\II{{\bf I}}
   To that goal, we define a structure \DDS{} which is illustrated by Figure~\ref{f_discr}. Each 
register is dotted with such a structure.

\vskip 10pt
{\bf The \DDS-structure}
\vskip 5pt

   There are two parts in the structure: one for the arrival of the locomotive from the program,
the other for the return of the locomotive when it comes back from the register, once its 
operation is completed. From the program, the locomotive is sent on a specific track attached to
the instruction it has to perform. Before the arrival to the register, the locomotive crosses
first a structure \DDI, \DDD{} if it has to increment, to decrement the register respectively.
That structure we describe further remembers which instruction had to be performed. That information
is recovered by the locomotive when it goes back from the register, which allows it to arrive at the
appropriate point of the program. Between \DDI{} or \DDD{} and the register, the locomotive crosses
\DDS{} which remembers the type of the instruction: to increment the register or to decrement it.

   The description of \DDS{} allows us to introduce various features we use in the description of
\DDI{} and of \DDD. In particular, we use different colours for the tracks in order to distinguish 
those which the locomotive runs when it has to increment the register from those it runs when it has 
to decrement it.

\vskip 10pt
\vtop{
\ligne{\hfill
\includegraphics[scale=0.5]{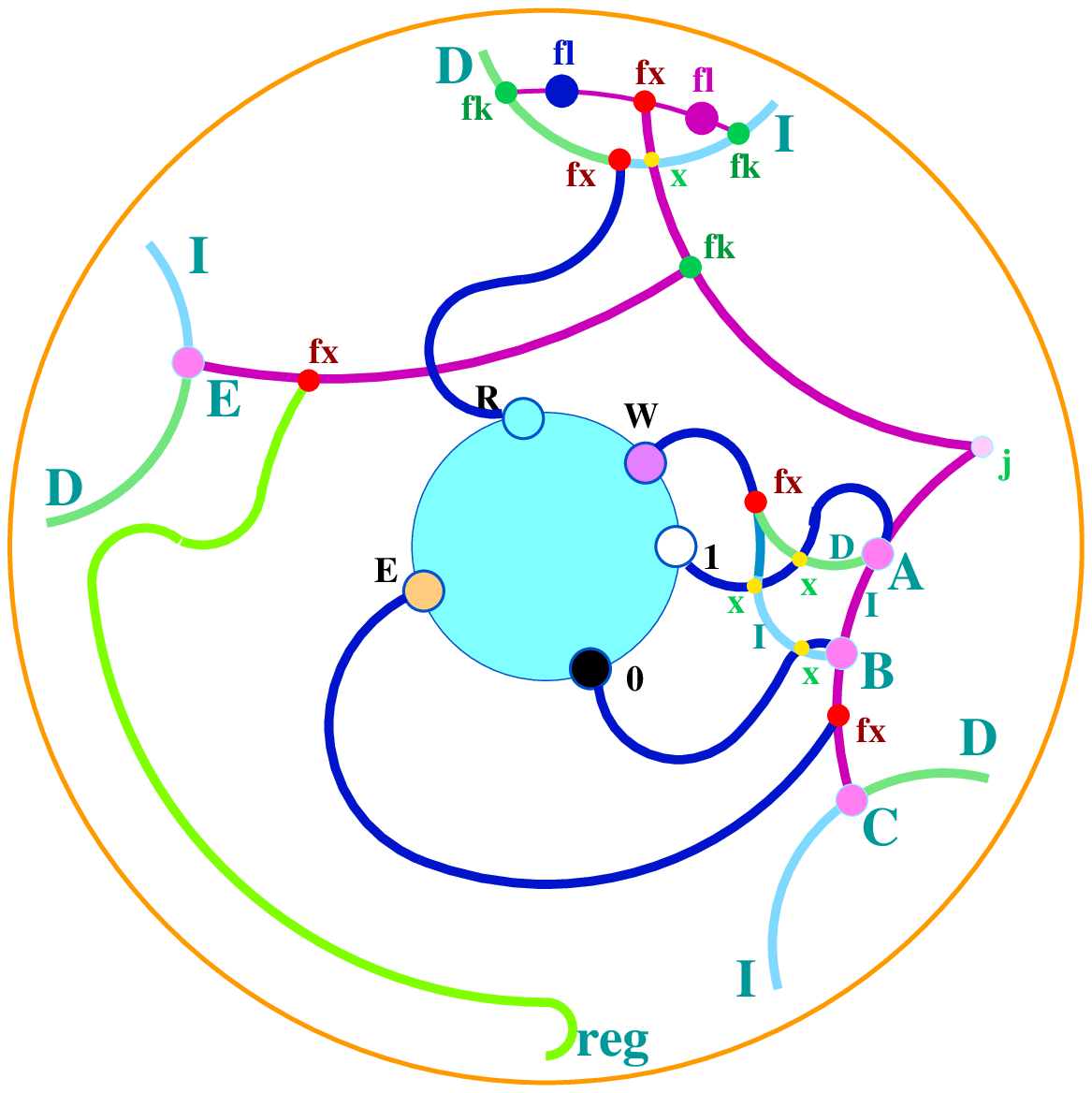}
\hfill}
\vspace{-30pt}
\begin{fig}\label{f_discr}
\leurre
The structure which memorises which type of instruction is sent to the register. Note the mauve 
tracks dispatching the information about the type of the instruction to all active memory switches:
at~$A$, at~$B$, at~$C$ and at~$E$. The filters sitting close to the entry are also changed but the 
corresponding mechanism is not illustrated in order to keep the figure as readable as possible.
Note that the track from $A$, $B$ to~\WW{} crosses the track from~\uu, \zz{} to~$A$, $B$ 
respectively.
\end{fig}
}

   When the locomotive arrives to \DDS, it enters through the track marked with~\II{} or with~\DD{}
respectively, lighter blue, lighter green for an incrementing, a decrementing instruction  
respectively. On the track, the locomotive meets a fork which sends it to a fixed switch whose
exit track leads to the \RR-entry of a one-bit memory. The fork sends a copy of the locomotive to
a filter. At initial time, the filter is mauve, blue on the side of the \II, \DD-track respectively.
The information possibly delivered by the filters is gathered by the fixed switch which sits on a 
track joining both filters. That possible information is conveyed through the mauve tracks of the
picture to be sent to four active memory switches displayed by the figure. At initial time, 
all those active memory switches select the track which the locomotive must follow when it has to
increment the register. 

   Consider the case when the locomotive has to increment the register. The locomotive arrives 
at the \RR-gate of the one-bit memory~$M$ sitting on the middle of the figure. The locomotive 
arrives at~\RR{} through the blue track of the figure, blue on the figure, the colour telling us
that such a track is used in both cases, whether to increment or to decrement the register. 
If it leaves~$M$ through~\uu, it is the confirmation that the the locomotive has to increment the 
register and so, it goes to the active memory switch sitting at~$A$ which to its turn send the
locomotive to~$C$. If it leaves the one-bit memory through~\zz, it means that the previous 
instruction performed on that register was to decrement it. The locomotive then goes from~\zz{} 
to~$B$. At~$B$, as far as the locomotive has to increment the register, the switch selects the 
track leading from~$B$ to the \WWW-gate. 
Crossing~$\mathcal M$ through~\WWW{} means that the locomotive rewrites the bit contained 
in~$\mathcal M$ from \zz{} to~\uu. Leaving~$\mathcal M$ through its \EE-gate, the locomotive goes 
to~$C$ through a fixed switch sitting on the segment of line~$\sigma$ joining the point {\bf j} 
to~$C$. Now, the active memory switch at~$C$ selects the \II-track, so that the locomotive later 
enters the register. If the locomotive leaves~$\mathcal M$ through~\uu, it is led to~$A$ where the 
active memory switch sends the locomotive to $\sigma$ so that it arrives at~$C$ where it goes on 
the \II-track.

   Note that at~$C$, even if the previous instruction was to decrement the register, the selected
track is the \II-one. If the previous information was to decrement the register, the filter closer 
to the \II-track entering the structure is blue, so that the information is sent to change the 
filters at the passive memory switch and at the active memory switches at~$A$, $B$, $C$ and~$E$. 
The information is conveyed by an auxiliary mauve locomotive running on the mauve tracks of the
figure. The particularity of the tracks followed by the mauve locomotive is that they consists of
segments of line, while the tracks followed by a blue locomotive, the tracks of the figure in a 
different colour, consists of arcs of circles. The difference is important: the segments of line 
run by the mauve locomotive are shorter than the arcs of circles followed by the blue locomotive. 
Accordingly, the signal sent to~$A$, $B$, $C$ and~$E$ reaches those switches before a blue 
locomotive crosses any of them.

   Presently, consider the case when the locomotive has to perform an instruction which
decrements the register. If the previous instruction was to increment the register, the
locomotive, after entering~$M$ through its \RR-gate, leaves~$M$ through~\uu. It then reaches~$A$
where the memory switch already selects the track from~$A$ to~\WWW. Accordingly, the locomotive
changes the bit contained in~$M$ from~\uu{} to~\zz. Leaving~$M$ through~\EE, the locomotive
joins~$C$ where it is sent to the \DD-track as far as the mauve locomotive sent from the entry 
arrived at~$C$ before the blue one does the same. Clearly, the fixed switch sitting between~$B$ 
and~$C$ is crossed by a possible mauve locomotive before a blue locomotive arrives there. The same 
observation holds for the switches at~$A$ and at~$B$. If the previous instruction was to decrement
the register, no mauve locomotive was sent as far as the selected tracks at the switches sitting
at~$A$, at~$B$, at~$C$ and at~$E$ are the right tracks leading to the \DD-track from~$C$.

   When the locomotive returns from the register, it arrives to the switch sitting at~$E$. As far
as no other locomotive has visited \DDS{} while the locomotive under consideration performed its
instruction on the register, the switch at~$E$ selects the appropriate track leading either 
to~\DDI{} or to~\DDD.

   We adapt the structure described for~\DDS{} to the next ones we shall study where it is needed 
to discriminate between two situations.

   Now, we turn to the \DDI-structure.
\vskip 10pt
{\bf The \DDI-structure}
\vskip 5pt

   It is crossed by the locomotive before arriving to~\DDS{} when the locomotive goes to the
register and it is crossed after visiting \DDS{} which knows which type of instruction the 
locomotive has performed on the register. 

The structure consists of as many units as there are instructions incrementing that register, 
say $R$, in the program. Each unit is based on a one-bit memory $\mathcal M$ and 
Figure~\ref{f_dispinc} illustrates such a unit. In that figure, the arrival is illustrated by blue 
tracks while the way back from the register is illustrated by light purple and red tracks.
The working of~\DDI{} is the following. An 
instruction for incrementing $R$ is connected through a path to a specific unit of~\DDI. The path 
goes from the program to the \WWW-gate of that unit. At the initial time, the configuration 
of~\DDI{} is such that all its unit contain~\zz: the switches of the one-bit memory 
are in a position which, by definition defines bit~\zz. Accordingly, when the locomotive 
enters the unit, it will change the flip-flop and the memory switches so that, 
by definition, the memory contains bit~\uu. The locomotive leaves the memory through 
the gate~\EE{} and it meets a flip-flop switch at~$A$ which, in its initial position, 
sends the locomotive to~$R$. Note that when the locomotive leaves~\DDI{} a single unit
of the structure contains the bit~\uu{} and selects a path to the program at the flip-flop
switch at~$A$. See the blue track of Figure~\ref{f_dispinc}.

\vskip 10pt
\ligne{\hfill
\vtop{
\ligne{\hfill
\includegraphics[scale=0.45]{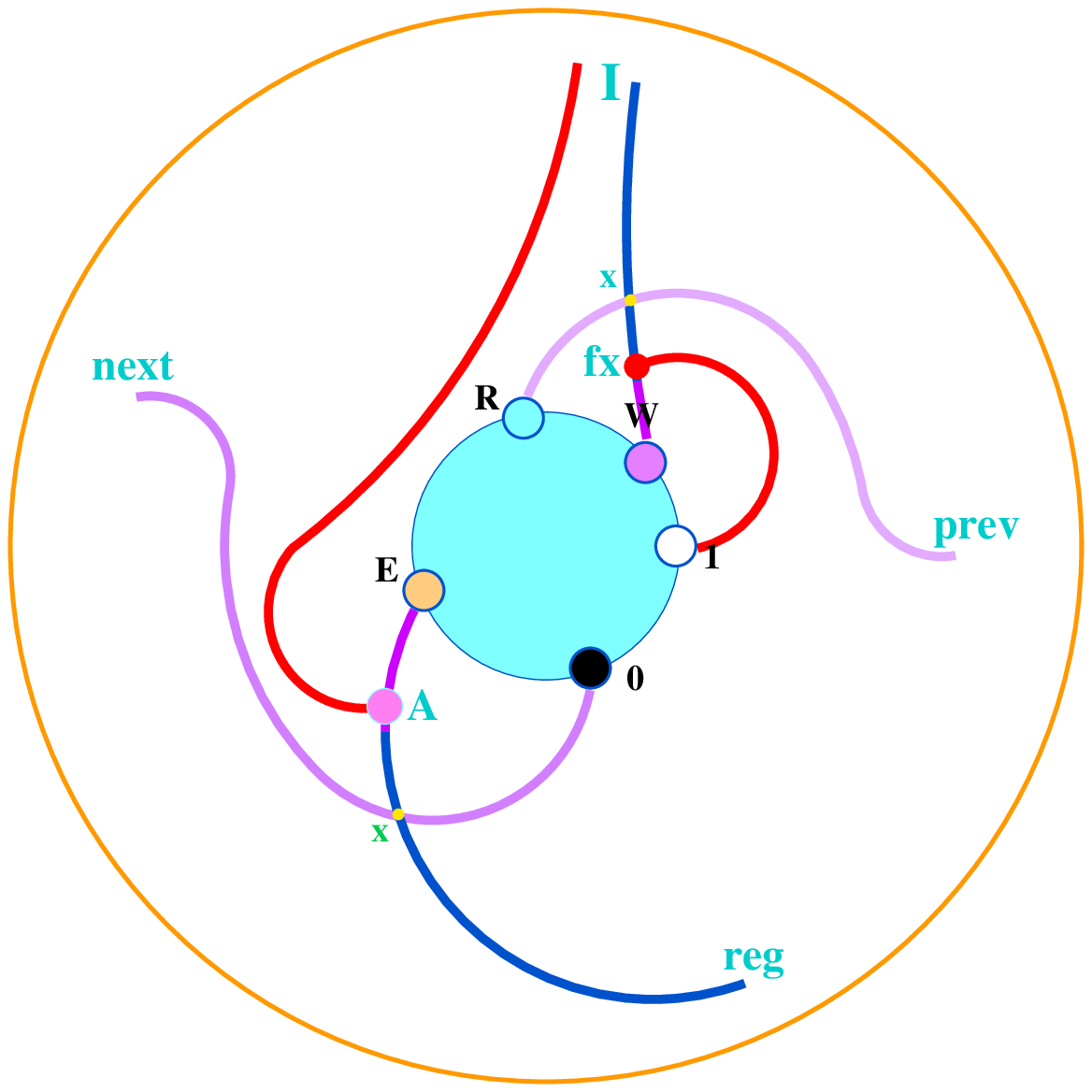}
\hfill}
\vspace{-10pt}
\begin{fig}\label{f_dispinc}
\leurre
The configuration of a unit of the structure which memorises the right incrementing
instruction. Note the three crossings in the implementation, note the colours of the tracks 
explained in the text. In dark purple, we have the tracks which are followed by the locomotive 
both when it arrives from the program, blue tracks, and when it goes from the \zz-gate to the 
\WWW-one, red tracks.
\end{fig}
}
\hfill}
\vskip 5pt
When the locomotive goes back from the register after it performed its operation, it goes back to
\DDS{} which knows that it incremented the register so that the locomotive is sent to
to \DDI. The locomotive reads the bit stored in $\mathcal M$. If it is~\zz, the
exit through the \zz-gate of~$\mathcal M$ leads the locomotive to the next unit, see the light
purple track of Figure~\ref{f_dispinc}. If it reads \uu, it knows that it reached the appropriate 
unit. It rewrites the unit of $\mathcal M$ as far as \uu{} sends the locomotive to the \WWW-gate 
of $\mathcal M$. From \EE, the locomotive joins $A$ where the flip-flop sends it back to the 
program, see the red tracks of Figure~\ref{f_dispinc}. As $A$ passed by the locomotive, the 
flip-flop sitting there selects the track to the register. Accordingly, the unit is ready for a 
new possible visit for the next instruction incrementing the register, possibly before operating 
on it or after the operation is completed.
%\newpage
\vskip 10pt
{\bf The \DDD-structure}
\vskip 5pt

Similarly, that structure memorises which instruction required to decrement that register. The 
structure contains as many unit as there are instructions of the program which decrements that 
register. The unit also contains a single one-bit memory which is~\zz{} before a locomotive arrives
at the structure. The locomotive enters the structure before visiting \DDS{} on its way to the
register and it visits the structure after leaving~\DDS{} which remembers that the locomotive has
decremented the register.

\vskip 10pt
\vtop{
\ligne{\hfill
\includegraphics[scale=0.5]{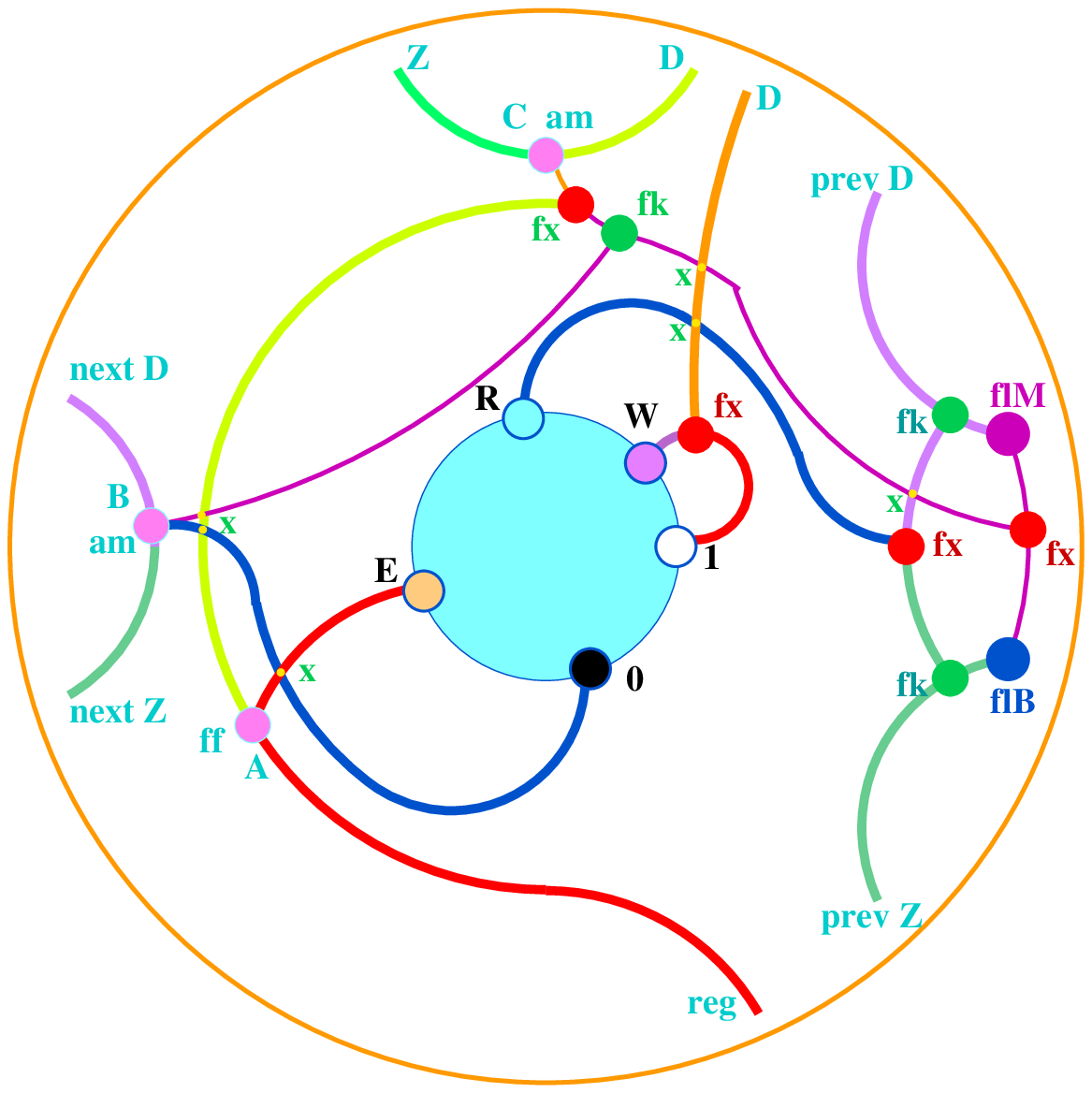}
\hfill}
\vspace{-30pt}
\begin{fig}\label{f_dispdec}
\leurre
The idle configuration of a unit of the structure which memorises the right decrementing
instruction. Note that the structure is more complex than that of Figure~{\rm\ref{f_dispinc}}.
Note the sketchy representation of the one-bit memory. Note the different colours of the tracks.
\end{fig}
}
\vskip 5pt
   We can see on Figure~\ref{f_dispdec} that a unit of~\DDD{} is more complex than that of~\DDI.
Indeed, when a locomotive arrives at a register to decrement it, it may happen that it could not
perform the operation because the register was empty, {\it i.e.} its first unit already 
contains~\zz. In that case, the locomotive leaves the register through a special track called the 
\ZZ-track which does not cross the~\DDS{} attached to the register. The \ZZ-track goes directly to 
the \DDD-structure. The track leaving \DDS{} as a former decrementing instruction is called a 
\DD-track and it arrives to the \DDD-structure at the same unit as the \ZZ-track. Accordingly, in 
each unit of~\DDD, it is important to know which track entered the unit after the instruction was 
performed. It is the reason of the filters displayed on the right-hand side border of the figure.
Those filters represent a passive memory switch which allows the unit to discriminate between 
the arrival from a \ZZ- or a \DD-track.

   In Figure~\ref{f_dispdec}  we also use different colours in order to illustrate the different 
ways taken by the locomotive. The arriving locomotive from the program is on an orange track and 
it arrives at a fixed switch whose exit track leads to the \WWW-gate of the unit.
As far as before the arrival of the locomotive the bit of the one-bit memory~$\mathcal M$ of the 
unit is~\zz, the bit has to be rewritten to~\uu, so that the locomotive arrives from the program 
to the \WWW-gate of the~$\mathcal M$ of that unit. Leaving~\EE, the locomotive is sent to a 
flip-flop switch {\bf ff} at $A$, see the red track on the figure, and {\bf ff} selects the track 
leading to the register. After the crossing of $A$, the flip-flop switch selects the other track 
which goes to the active memory switch selecting the return \ZZ- or \DD-track to the program. 

   When the locomotive comes back from the register, it arrives to ~\DDD{} either through 
a~\ZZ-track, green track on the figure, or through a \DD-one, purple track on the figure. The 
filters displayed on the figure represent a passive memory switch. The mauve locomotive possibly 
sent possibly changes the selection of the active memory switches in order they select the track 
corresponding to the one observed by the passive memory switch. If the selection corresponds to 
the arrival track, no change occurs but if it is not the case the change is performed at the 
passive switch as well as the active memory switches reached by the mauve locomotive before any 
blue locomotive arrives at the same switches. 

Returning from the register either through the \ZZ-track or through the \DD-one, the locomotive 
arrives to the \RR-gate of~$M$ through a common blue track on the figure. If it reads \zz, it goes 
to the next unit: from~\zz, it goes to~$B$, see the blue track from~\zz{} to $B$. At $B$ the
active memory switch sends the locomotive to the next unit through the \DD- or the \ZZ-track,
purple or green tracks on the figure, respectively.
%crossing the track from \EE{} to~$A$ and crossing a fixed switch whose exit track leads to~$B$ 
%where the active memory switch selects the appropriate track to the next unit. 
If the locomotive reads~\uu, it knows that it is at the unit which allows to return to the right 
place of the program. But before going there, the locomotive has to rewrite the bit 
in~$\mathcal M$ from~\uu{} to \zz. It is why there is a track from the~\uu-gate joining it 
with \WWW, see the red track on the figure. To that purpose the track joins 
that from the program to~\WWW{} thanks to a fixed switch whose exit track leads to~\WWW. The 
locomotive leaves~$\mathcal M$ through \EE{} from where the track leads it to~$A$ where a 
flip-flop is sitting, selecting a track to~$C$. Once the flip-flop at~$A$ is crossed, it selects 
again the track to the register so that at that moment, the locomotive may leave the unit which is 
in its \zz-state. When it is at~$C$, the active memory switch sitting there has already be informed
whether the instruction arrived from a \ZZ- or a \DD-track. Accordingly, the locomotive is sent to 
the right place of the program: either to the instruction which stands after the just executed one 
or, if the \ZZ-track was used, to a particular instruction, performing that way the execution of a 
jump instruction. The \DD- or \ZZ-track is on light green or darker green respectively on 
Figure~\ref{f_dispdec}.

\subsubsection{Constitution of a register and operating upon it}\label{sssreg}

   The implementation of the register requires a special examination. Weak universality
means that the initial configuration is infinite but not arbitrary. In the present paper,
it will be periodic outside a large ball containing the implementation part of the
program and also the first unit of the two registers needed for universality, according
to Minsky's theorem, see~\cite{minsky}. Each register, in some sense, follows a line
and that construction along each line is periodic, being infinite in one direction.

   A register consists of infinitely many units which we may index by $\mathbb N$.
Let $\mathcal R$ denote a register. By $\mathcal R$($n$), we denote the $n^{\rm th}$
unit. We shall call $\mathcal R$(0) the first unit of the register. Each unit contains a one-bit 
memory~$\mathcal M$. The memory contains \zz{} or~\uu. At each time~$t$ of the computation, there is
a number~$c_t$ such that the bit of $\mathcal R$($n$) is always set to~\zz{} when
\hbox{$n\geq c_t$} and all of them are set to~\uu{} when \hbox{$n<c_t$}. We say that
$c_t$ is the {\bf value} of the register. We also say that it is its {\bf content}.
When $c_t=0$ we also say that the register is empty. In that case, the bit in the
memory of every unit of the register is set to~\zz. To left, a passive memory switch recognizes 
the information telling whether the operation to perform is to increment $\mathcal R$ or to 
decrement it. That information is transferred to the active memory switches $S0$, $S1$ and $So$ in 
order to appropriately select the tracks. An incrementing instruction changes the first \zz-bit 
which it meets with \uu{} and it goes back to \DDS. A decrementing instruction detects the first 
\zz-bit which it meets and it goes back to the previous unit where it changes the \uu-bit to \zz{} 
and then it goes back to \DDS. Figure~\ref{f_unitreg} illustrates all those workings including the 
case when $\mathcal R$ is empty. In that case, the return track to the previous unit is replaced by
the initial part of the \ZZ-track.

%\vskip 10pt
\vtop{
\ligne{\hfill
\includegraphics[scale=0.5]{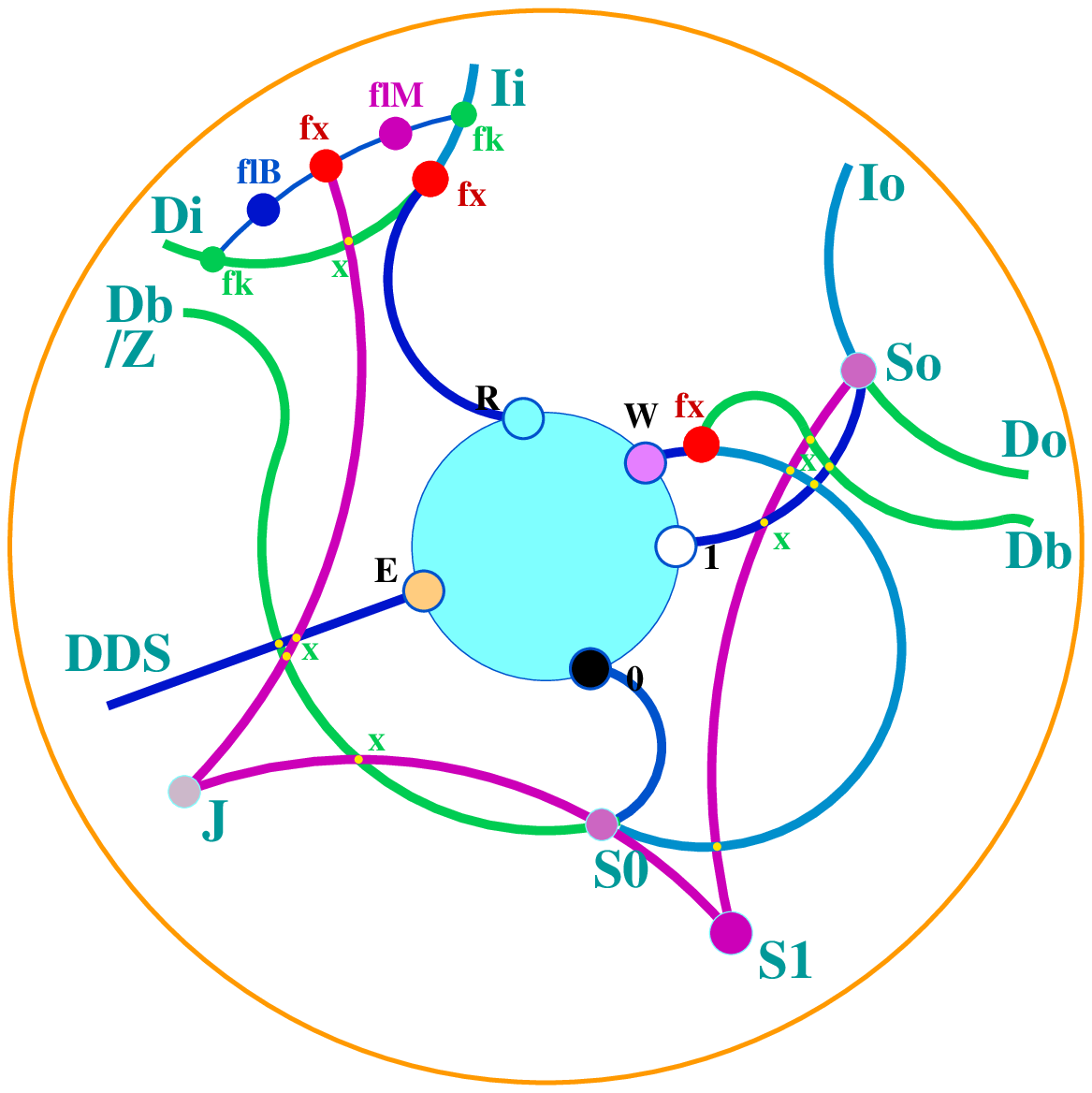}
\hfill}
\vspace{-30pt}
\begin{fig}\label{f_unitreg}
\leurre
The idle configuration of a unit of a register. We can see the one-bit memory and the memory
switches devoted to the materialisation of the content of the register. The {\bf Db$/$Z} mention
of the figure indicates that in the first unit if the register that track is the beginning of the
\ZZ-track. On the other units, {\bf Db} indicates the track leading to the previous unit of the
register.
\end{fig}
}

   Let us describe the working of an incrementing instruction. Figure~\ref{f_unitreg} represents a 
unit of the register, colouring the tracks in the same way as in the previous figures. The 
locomotive arrives through the {\bf Ii}-track, see the figure. The locomotive arrives to \RR. If 
it reads \uu, leaving $\mathcal M$ through its \uu-gate, it goes to~$So$ where the active memory 
switch sends it towards the next unit on the \II-track. If it reads \zz, the locomotive knows 
that it is the first time it meets \zz{} since its arrival to the register. Accordingly, it goes 
to~$S0$ too where the active memory switch sends it to \WWW. It rewrites the bit of~$\mathcal M$ 
from~\zz{} to \uu{} and then it goes to~\EE{} to join \DDS.

   Presently consider the working of a decrementing instruction. The locomotive arrives through the
{\bf Di}-track, see Figure~\ref{f_unitreg}. The locomotive again arrives to~\RR. If it reads \uu, it
leaves $\mathcal M$ through its \uu-gate, then it arrives at $So$ where the active memory switch 
sends it onto the \DD-track towards the next unit. If the locomotive reads \zz{} it knows that it 
reached the first cell after the \uu's realising the value of the register. So that the locomotive
must rewrite the \uu-bit of the previous unit to~\zz. Accordingly, going to $S0$, the active memory 
switch sends it to the track {\bf Db} which goes to the previous unit. In the previous unit, using 
the same picture of Figure~\ref{f_unitreg}, we can see that the {\bf Db}-track arrives at the 
\WWW-gate of the $\mathcal M$ of that unit through a fixed switch whose exiting track goes to \WWW. 
The locomotive exit through~\EE, running on the track which leads to \DDS. So that when an 
operation is performed the locomotive always return through a single track going to \DDS. There is 
an exceptional case: when $\mathcal R$ is empty. In the case of a decrementing instruction, the 
locomotive reads~\zz, so that in that case, the {\bf Db}-track is nothing else than the \ZZ-track 
going into \DDD. 

\section{The table}\label{stable}

   From the definition of a weighted cellular automaton, the transition function is defined
by two data: the current state of the cell and the neighbouring weight of that cell. Accordingly,
as already mentioned, the input and the output defining the transition can be put in form of a
table. The first entry is a state, the second one is a non negative integer.

   In order to explain that table, we refer to a table which gathers the maximal amount of 
information. Table~\ref{t_small} assembles the whole transition function as a table whose columns
are headed by explaining labels: 'n°' stands for the number of the entry, 
{\footnotesize{c-YBRMV:n}} displays the neighbourhood and the function: 'c' stands for the current 
state, 'n' stands for the new one and under each letter representing a state~$\sigma$, the number 
of neighbours under~$\sigma$; at last and not the least '$\Sigma$' stands for the neighbouring
weight. The representation of the neighbourhood allows us to know the neighbourhood giving rise
to the corresponding weight and also to know how the sum can be decomposed into the weights of
the neighbouring states. Table~\ref{t_small} gives the above information in the case when both the
current state and the new one are not identical to \WW.

   Tables~\ref{t_tracks}  is an extract of Table~\ref{t_small} devoted to the running of a 
locomotive over tracks an over forks. We can see that the entries of Table~\ref{t_small} split into 
two cases: when \hbox{'c' = 'n'} and when \hbox{'c' $\not=$ 'n'}. The first case can be decomposed
into two sub-cases: when the neighbourhood also does not change and when it does. In the first 
sub-case we speak of a conservative entry while in the second one, we speak of a witnessing entry. 
In the case when \hbox{'c' $\not=$ 'n'} we speak of a motion entry. In most cases, motion entries 
deal with cells belonging to the tracks or which are changed by the close occurrence of the 
locomotive, witnessing entries deal with cells which can see the locomotive during its motion 
while conservative entries deal with the cells which, together with their neighbourhood, are never 
changed or by cells which belong to an idle configuration.

Most often, motion entries obey the following pattern:
\vskip 10pt
\ligne{\hfill
$\vcenter{\vtop{\leftskip 0pt\parindent 0pt\hsize=150pt
\ligne{\hfill\WW,$v$+4 $\rightarrow$ \BB,\hfill}
\ligne{\hfill\BB,$v$+12 $\rightarrow$ \RR,\hfill}
\ligne{\hfill\RR,$v$+4 $\rightarrow$ \WW,\hfill}
\ligne{\hfill\WW,$v$+12 $\rightarrow$ \WW\hfill}
}}$
\hskip 20pt
$\vcenter{\vtop{\leftskip 0pt\parindent 0pt\hsize=150pt
\ligne{\hfill\WW,$v$+29 $\rightarrow$ \MM,\hfill}
\ligne{\hfill\MM,$v$+12 $\rightarrow$ \RR,\hfill}
\ligne{\hfill\RR,$v$+29 $\rightarrow$ \WW,\hfill}
\ligne{\hfill\WW,$v$+12 $\rightarrow$ \WW\hfill}
}}$
\hfill(\numerrel)\hskip 10pt}
\vskip 5pt
\noindent
For the witness cells the entries obey a similar pattern as (4), except that the state is not 
changed. There is an additional pattern: 
\vskip 5pt
\ligne{\hfill$\eta$,$v$+16 $\rightarrow$ $\eta$, \hskip 20pt
$\eta$,$v$+41 $\rightarrow$ $\eta$\hfill (\numerrel)\hskip 10pt}
\vskip 5pt
\noindent
for a blue, mauve locomotive respectively. Those additional formulas are used by 
cells that can be neighbours of both cells of a locomotive. 

Tables~\ref{t_tracks} up to~\ref{t_prog} display the entries as above mentioned together with
tables indicating which entries apply to a few selected cells. Later, a condensed table
giving the new state for a couple consisting of a state and a neighbourhood weight is given: 
Table~\ref{t_small}. Those tables prove the number of entries for the table mentioned in
Theorem~\ref{letheo}. The number given in the present paper much improves that given 
in~\cite{mmarXiv23c}. In the present paper, the table does not contain repetitions. If a 
neighbouring occurs several time, each occurrence happens with a different current state.
%of~\cite{mmarXiv23b} was cleaned as far as 
%that table contained several repetitions.

We illustrate those tables by application of the considered lines to specific cells. Each cell is 
taken from a figure of the paper. We indicate times starting from~0, a time at which the cell is
idle. We also indicate the evolution in time of the state of the cell together with the number
of the entry of Table~\ref{t_small} which applies to the cell at that time. We also indicate
the neighbouring weight for the corresponding entries.

We start with cell (3,2) of Figure~\ref{f_rac_i}. The application of Tables~\ref{t_tracks},
or \ref{t_small} is to be find in lines (tr b), (tr m) for a blue, mauve locomotive respectively.

\newdimen\tttnca\tttnca=15pt
\newdimen\tttncb\tttncb=60pt

\def\tttncourt #1 #2 #3 {%
\hbox to \tttnca {\hfill #1\hfill}
\hbox to \tttncb {\hfill #2\hfill}
\hbox to \tttnca {\hfill #3\hfill}
}

\newdimen\tttncc\tttncc=25pt
\def\tttlcourt #1 #2 #3 #4 #5 {%
\hbox to \tttnca {\hfill #1\hfill}
\hbox to \tttncb {\hfill #2\hfill}
\hbox to \tttnca {\hfill #3\hfill}
\hbox to \tttncc {\hfill #4\hfill}
\hbox to \tttnca {\hfill #5\hfill}
}

\newdimen\gtablarge\gtablarge=25pt 

\newdimen\stablarge\stablarge=19pt
\def\tttable #1 #2 #3 #4 #5 #6 #7 #8 #9 {%
\hbox to 200pt{%
\hbox to \gtablarge{\hfill#1\hfill}
\hbox to \gtablarge{\hfill#2\hfill}
\hbox to \stablarge{\hfill#3\hfill}
\hbox to \stablarge{\hfill#4\hfill}
\hbox to \stablarge{\hfill#5\hfill}
\hbox to \stablarge{\hfill#6\hfill}
\hbox to \stablarge{\hfill#7\hfill}
\hbox to \stablarge{\hfill#8\hfill}
\hbox to \stablarge{\hfill#9\hfill}
}
}
\def\adtttable #1 #2 #3 #4 {%
\hbox to 120pt{%
\hbox to \gtablarge{\hfill#1\hfill}
\hbox to \gtablarge{\hfill#2\hfill}
\hbox to \gtablarge{\hfill#3\hfill}
\hbox to \gtablarge{\hfill#4\hfill}
}
}

\newdimen\slarge\slarge=25pt
\def\tttbcds #1 #2 #3 #4 {%
\hbox to \slarge{\hfill#1}
\hbox to \slarge{\hfill#2}
\hbox to \slarge{\hfill#3}
\hbox to \slarge{\hfill#4}
}

\newdimen\rclarge\rclarge=25pt
\def\tttrc #1 #2 #3 #4 #5 #6 #7 #8 {%
\hbox to \rclarge{\hfill#1\hfill}
\hbox to \rclarge{\hfill#2\hfill}
\hbox to \rclarge{\hfill#3\hfill}
\hbox to \rclarge{\hfill#4\hfill}
\hbox to \rclarge{\hfill#5\hfill}
\hbox to \rclarge{\hfill#6\hfill}
\hbox to \rclarge{\hfill#7\hfill}
\hbox to \rclarge{\hfill#8\hfill}
}

\vskip 5pt
\ligne{\hfill
$\vcenter{\hbox{\vtop{\leftskip 0pt\parindent 0pt\hsize=250pt
\ligne{\hfill\hbox to \rclarge{\hfill time\hfill} 
\tttrc 0       1       2       3       4       5       6       7        {} \hfill}
%{(2,1)} {(3,1)} {(2,4)} {(3,2)} {(3,3)} {(3,4)} {(3,5)} {(3,10)} {} \hfill}
\ligne{\hfill\hbox to \rclarge{\hfill state\hfill}
\tttrc {\YY}   {\YY}   {\YY}   {\BB}   {\RR}   {\YY}   {\YY}   {\YY}    {} \hfill}
\ligne{\hfill\hbox to \rclarge{\hfill sum\hfill}
\tttrc {2}    {2}    {5}    {13}   {5}    {13}    {2}    {2}     {} \hfill}
\ligne{\hfill\hbox to \rclarge{\hfill line\hfill}
\tttrc {1}    {1}    {3}    {4}    {5}    {6}    {1}    {1}     {} \hfill}
}}}$
\hfill (tr b)\hskip 15pt}

\vskip 10pt
\ligne{\hfill
$\vcenter{\hbox{\vtop{\leftskip 0pt\parindent 0pt\hsize=250pt
\ligne{\hfill\hbox to \rclarge{\hfill time\hfill} 
\tttrc 0       1       2       3       4       5       6       7        {} \hfill}
%{(2,1)} {(3,1)} {(2,4)} {(3,2)} {(3,3)} {(3,4)} {(3,5)} {(3,10)} {} \hfill}
\ligne{\hfill\hbox to \rclarge{\hfill state\hfill}
\tttrc {\YY}   {\YY}   {\YY}   {\MM}   {\RR}   {\YY}   {\YY}   {\YY}    {} \hfill}
\ligne{\hfill\hbox to \rclarge{\hfill sum\hfill}
\tttrc {2}    {2}    {30}    {13}    {30}    {13}    {2}    {2}     {} \hfill}
\ligne{\hfill\hbox to \rclarge{\hfill line\hfill}
\tttrc {1}    {1}    {8}    {9}    {10}    {6}    {1}    {1}     {} \hfill}
}}}$
\hfill (tr m)\hskip 15pt}
\vskip 5pt
In those lines, we can see that the front of the locomotive is seen in the cell~$c$$-$1 at time~2, 
that the front of the locomotive is in the cell~$c$ at time~3, that its rear is at that cell at
time~4 while the front is at the cell $c$$+$1 and that, at time~5 the state of the cell~$c$ is
again \YY{} while the rear of the locomotive is seen in the cell~$c$$+$1.

\ifnum 1=0 {
\ligne{\hfill
\vtop{\leftskip 0pt\parindent 0pt\hsize=220pt
\begin{tab}\label{t_trc}
\leurre
The motion of a locomotive along a track. The first line indicates the cell under consideration.
The second line indicate their states when the configuration is idle.
\end{tab}
\ligne{\tttrc {(2,1)} {(3,1)} {(2,4)} {(3,2)} {(3,3)} {(3,4)} {(3,5)} {(3,10)} {} \hfill}
\ligne{\tttrc {\GG}   {\GG}   {\WW}   {\WW}   {\WW}   {\WW}   {\YY}   {\YY}    {} \hfill}
\ligne{\tttrc {45}    {45}    {17}    {24}    {17}    {17}    {13}    {13}     {} \hfill}
\ligne{\tttrc {46}    {45}    {21}    {24}    {17}    {17}    {13}    {13}     {} \hfill}
\ligne{\tttrc {47}    {45}    {22}    {25}    {17}    {17}    {14}    {13}     {} \hfill}
\ligne{\tttrc {47}    {46}    {23}    {26}    {21}    {17}    {16}    {13}     {} \hfill}
\ligne{\tttrc {34}    {47}    {d.}    {27}    {22}    {21}    {15}    {14}     {} \hfill}
\ligne{\tttrc {45}    {47}    {17}    {d.}    {23}    {22}    {13}    {16}     {} \hfill}
\ligne{\tttrc {45}    {47}    {17}    {24}    {d.}    {23}    {13}    {15}     {} \hfill}
\ligne{\tttrc {45}    {45}    {17}    {24}    {17}    {17}    {13}    {13}     {} \hfill}
}
\hfill}
\vskip 10pt
} \fi

Comparing lines (tr b) and (tr m), we can see the difference of the neighbouring sum entailed by
the difference of weight between \BB{} and \MM, 4 and 29 respectively. At time~5, the rear is seen 
from $c$, when the same neighbouring sum and the same entry applies. Note that entries~13 and~17 
have the same neighbourhood {\footnotesize {20100}} and the same neighbourhood weight 14 but
the current state is different.

\ifnum 1=0 {
Tables~\ref{t_tracks_a} and \ref{t_tracks_b} also contain the lines applying to a passive fixed 
switch and to a fork. Consider what happens at cell~0 of Figure~\ref{f_fix_i}, see lines~(fx b)
and~(fx m).

On those tables, we can see that the neighbourhood weights of cell~0 are
different from those of
cell (3,2) illustrated by lines (tr b) and (tr m). Tables~\ref{t_tracks_a} and \ref{t_tracks_b}
explain that difference as illustrated by Figures~\ref{f_rac_i} and \ref{f_fix_i} respectively.
Note that at time~3, line 56 applies instead of line 26 as far as shown by Figure~\ref{f_fix_m}
and Table~\ref{t_tracks_b} two \YY-neighbours occur instead of a single one for (3,2) of
Figure~\ref{f_rac_i}.

As mentioned, Tables~\ref{t_tracks_a} and~\ref{t_tracks_b} also apply to a fork. We 
give them in lines (fk b) and (fk m), those latter ones being displayed after 
Table~\ref{t_tracks_a}.
} \fi

\vskip 5pt
\ligne{\hfill
$\vcenter{\hbox{\vtop{\leftskip 0pt\parindent 0pt\hsize=250pt
\ligne{\hfill\hbox to \rclarge{\hfill time\hfill} 
\tttrc 0       1       2       3       4       5       6       7        {} \hfill}
\ligne{\hfill\hbox to \rclarge{\hfill state\hfill}
\tttrc {\YY}   {\YY}   {\YY}   {\BB}   {\RR}   {\YY}   {\YY}   {\YY}    {} \hfill}
\ligne{\hfill\hbox to \rclarge{\hfill sum\hfill}
\tttrc {3}    {3}    {6}    {14}    {9}   {25}    {3}    {3}     {} \hfill}
\ligne{\hfill\hbox to \rclarge{\hfill line\hfill}
\tttrc {11}    {11}    {12}    {13}    {14}   {15}    {11}    {11}     {} \hfill}
}}}$
\hfill (fk b)\hskip 15pt}

\vskip 10pt
\ligne{\hfill
$\vcenter{\hbox{\vtop{\leftskip 0pt\parindent 0pt\hsize=250pt
\ligne{\hfill\hbox to \rclarge{\hfill time\hfill} 
\tttrc 0       1       2       3       4       5       6       7        {} \hfill}
\ligne{\hfill\hbox to \rclarge{\hfill state\hfill}
\tttrc {\YY}   {\YY}   {\YY}   {\MM}   {\RR}   {\YY}   {\YY}   {\YY}    {} \hfill}
\ligne{\hfill\hbox to \rclarge{\hfill sum\hfill}
\tttrc {3}    {3}    {31}    {14}    {59}   {25}    {3}    {3}     {} \hfill}
\ligne{\hfill\hbox to \rclarge{\hfill line\hfill}
\tttrc {11}    {11}    {16}    {17}    {18}   {15}    {11}    {11}     {} \hfill}
}}}$
\hfill (fk m)\hskip 15pt}
\vskip 5pt

\vskip 10pt
\ligne{\hfill
\vtop{
\begin{tab}\label{t_tracks}
\leurre
Table explaining the control table for the idle configurations of the tracks, also across the 
fork.  The lines also deal with the motion of the locomotive both for a blue and a mauve one.
\end{tab}
\ligne{\hfill
\vtop{\leftskip 0pt\parindent 0pt\hsize=120pt
\ligne{\hfill\tttlcourt {n°} {{\footnotesize{c-YBRMV:n}}} {$\Sigma$} {c.} {F.} \hfill}
\vskip 3pt
\ligne{\hfill tracks\hfill}
\ligne{\hfill\tttlcourt {  0} {0-00000:0} {0} {0} {\ref{f_rac_m}} \hfill}
\ligne{\hfill\tttlcourt {  1} {1-20000:1} {2} {(1,5)} {id.} \hfill}
\ligne{\hfill\tttlcourt {  2} {1-10000:1} {1} {(1,14)} {id.} \hfill}
\ligne{\hfill blue locomotive\hfill}
\ligne{\hfill\tttlcourt {  3} {1-11000:2} {5} {(7,8)} {id.} \hfill}
\ligne{\hfill\tttlcourt {  4} {2-10100:3} {13} {(7,7)} {id.} \hfill}
\ligne{\hfill\tttlcourt {  5} {3-11000:1} {5} {(7,6)} {id.} \hfill}
\ligne{\hfill\tttlcourt {  6} {1-10100:1} {13} {(7,7)} {id.} \hfill}
\ligne{\hfill\tttlcourt {  7} {1-00100:1} {12} {(7,5)} {id.} \hfill}
\ligne{\hfill mauve locomotive\hfill}
\ligne{\hfill\tttlcourt {  8} {1-10010:4} {30} {(7,8)} {id.} \hfill}
\ligne{\hfill\tttlcourt {  9} {4-10100:3} {13} {(7,7)} {id.} \hfill}
}
\hfill
\vtop{\leftskip 0pt\parindent 0pt\hsize=100pt
\ligne{\hfill\tttlcourt {n°} {{\footnotesize{c-YBRMV:n}}} {$\Sigma$} {c.} {F.} \hfill}
\vskip 3pt
\ligne{\hfill\tttlcourt { 10} {3-10010:1} {30} {(7,6)} {\ref{f_rac_m}} \hfill}
\ligne{\hfill fork\hfill}
\ligne{\hfill\tttlcourt { 11} {1-30000:1} {3} {(4,1)} {\ref{f_frk_m}} \hfill}
\ligne{\hfill blue locomotive\hfill}
\ligne{\hfill\tttlcourt { 12} {1-21000:2} {6} {id.} {id.} \hfill}
\ligne{\hfill\tttlcourt { 13} {2-20100:3} {14} {id.} {id.} \hfill}
\ligne{\hfill\tttlcourt { 14} {3-12000:1} {9} {id.} {id.} \hfill}
\ligne{\hfill\tttlcourt { 15} {1-10200:1} {25} {id.} {id.} \hfill}
\ligne{\hfill mauve locomotive\hfill}
\ligne{\hfill\tttlcourt { 16} {1-20010:4} {31} {id.} {id.} \hfill}
\ligne{\hfill\tttlcourt { 17} {4-20100:3} {14} {id.} {id.} \hfill}
\ligne{\hfill\tttlcourt { 18} {3-10020:1} {59} {id.} {id.} \hfill}
}
\hfill}
}
\hfill}
\vskip 10pt

Entries 14 and 18 which appear at time~4{} in lines (fk b) and (fk m) mention the occurrences
of the front of two locomotives in the neighbourhood of the cell~(4,1) from Figures~\ref{f_auxil}
and~\ref{f_frk_m}.

Let us now look at the passive fixed switch. Table~\ref{t_fixed} displays the lines of
Table~\ref{t_small} which deal with the passive fixed switch. Here to the information given in
Table~\ref{t_small}, we indicate for each entry the cell where its application first appears.
Lines (fx b) and (fx m) show us which entry apply to cell~0 when a blue and a mauve locomotive
respectively is running across the switch. As can be seen, entries applying for a blue locomotive
may also apply to a mauve one. For instance, as the central cell remains most always \VV, the
same entry applies to that cell when it can see the rear of the locomotive in the cell (4,1) as far
as the rear is always \RR.

\vskip 5pt
\ligne{\hfill
$\vcenter{\hbox{\vtop{\leftskip 0pt\parindent 0pt\hsize=250pt
\ligne{\hfill\hbox to \rclarge{\hfill time\hfill} 
\tttrc 0       1       2       3       4       5       6       7        {} \hfill}
\ligne{\hfill\hbox to \rclarge{\hfill state\hfill}
\tttrc {\VV}   {\VV}   {\VV}   {\RR}   {\VV}   {\VV}   {\VV}   {\VV}    {} \hfill}
\ligne{\hfill\hbox to \rclarge{\hfill sum\hfill}
\tttrc {37}    {37}    {40}    {17}    {48}    {37}    {37}    {37}     {} \hfill}
\ligne{\hfill\hbox to \rclarge{\hfill line\hfill}
\tttrc {21}    {21}    {26}    {27}    {34}    {21}    {21}    {21}     {} \hfill}
}}}$
\hfill (fx b)\hskip 15pt}

\vskip 5pt
\ligne{\hfill
$\vcenter{\hbox{\vtop{\leftskip 0pt\parindent 0pt\hsize=250pt
\ligne{\hfill\hbox to \rclarge{\hfill time\hfill} 
\tttrc 0       1       2       3       4       5       6       7        {} \hfill}
\ligne{\hfill\hbox to \rclarge{\hfill state\hfill}
\tttrc {\VV}   {\VV}   {\RR}   {\VV}   {\VV}   {\VV}   {\VV}   {\VV}    {} \hfill}
\ligne{\hfill\hbox to \rclarge{\hfill sum\hfill}
\tttrc {37}    {37}    {65}    {76}    {48}    {37}    {37}    {37}     {} \hfill}
\ligne{\hfill\hbox to \rclarge{\hfill line\hfill}
\tttrc {21}    {21}    {39}    {43}    {34}    {21}    {21}    {21}     {} \hfill}
}}}$
\hfill (fx m)\hskip 15pt}
\vskip 10pt
\ligne{\hfill
\vtop{\leftskip 0pt\parindent 0pt
\begin{tab}\label{t_fixed}
\leurre
Table of the entries of Table~{\rm \ref{t_small}} which deal with the passive fixed switch.
\end{tab}
\vskip 0pt
\ligne{\hfill
\vtop{\leftskip 0pt\parindent 0pt\hsize=100pt
\ligne{\hfill\tttlcourt {n°} {{\footnotesize{c-YBRMV:n}}} {$\Sigma$} {c.} {F.} \hfill}
\vskip 3pt
\ligne{\hfill fixed switch\hfill}
\ligne{\hfill\tttlcourt { 19} {1-30001:1} {37} {(4,1)} {\ref{f_fix_i}} \hfill}
\ligne{\hfill\tttlcourt { 20} {1-20001:1} {36} {(5,1)} {id.} \hfill}
\ligne{\hfill\tttlcourt { 21} {5-30001:5} {37} {0} {id.} \hfill}
\ligne{\hfill\tttlcourt { 22} {5-00001:5} {34} {(1,1)} {id.} \hfill}
\ligne{\hfill blue locomotive\hfill}
\ligne{\hfill\tttlcourt { 23} {1-11001:2} {39} {(3,1)} {\ref{f_fix_m}} \hfill}
\ligne{\hfill\tttlcourt { 24} {2-10101:3} {47} {id.} {id.} \hfill}
\ligne{\hfill\tttlcourt { 25} {1-21001:2} {40} {(4,1)} {id.} \hfill}
\ligne{\hfill\tttlcourt { 26} {5-21001:3} {40} {0} {id.} \hfill}
\ligne{\hfill\tttlcourt { 27} {3-11100:1} {17} {(3,1)} {id.} \hfill}
\ligne{\hfill\tttlcourt { 28} {1-11100:1} {17} {(5,1)} {id.} \hfill}
\ligne{\hfill\tttlcourt { 29} {2-20200:3} {26} {(4,1)} {id.} \hfill}
\ligne{\hfill\tttlcourt { 30} {3-11101:5} {51} {0} {id.} \hfill}
\ligne{\hfill\tttlcourt { 31} {5-00100:5} {12} {(1,1)} {id.} \hfill}
}
\hfill
\vtop{\leftskip 0pt\parindent 0pt\hsize=100pt
\ligne{\hfill\tttlcourt {n°} {{\footnotesize{c-YBRMV:n}}} {$\Sigma$} {c.} {F.} \hfill}
\vskip 3pt
\ligne{\hfill\tttlcourt { 32} {3-21001:1} {40} {(4,1)} {\ref{f_fix_m}} \hfill}
\ligne{\hfill\tttlcourt { 33} {1-10101:1} {47} {(5,1)} {id.} \hfill}
\ligne{\hfill\tttlcourt { 34} {5-20101:5} {48} {0} {id.} \hfill}
\ligne{\hfill\tttlcourt { 35} {1-20101:1} {48} {(4,1)} {id.} \hfill}
\ligne{\hfill mauve locomotive\hfill}
\ligne{\hfill\tttlcourt { 36} {1-10011:4} {64} {(3,1)} {id.} \hfill}
\ligne{\hfill\tttlcourt { 37} {4-10101:3} {47} {(4,1)} {id.} \hfill}
\ligne{\hfill\tttlcourt { 38} {1-20011:4} {65} {id.} {id.} \hfill}
\ligne{\hfill\tttlcourt { 39} {5-20011:3} {65} {0} {id.} \hfill}
\ligne{\hfill\tttlcourt { 40} {3-10110:1} {42} {(3,1)} {id.} \hfill}
\ligne{\hfill\tttlcourt { 41} {1-10110:1} {42} {(5,1)} {id.} \hfill}
\ligne{\hfill\tttlcourt { 42} {4-20200:3} {26} {(4,1)} {id.} \hfill}
\ligne{\hfill\tttlcourt { 43} {3-10111:5} {76} {0} {id.} \hfill}
\ligne{\hfill\tttlcourt { 44} {3-20011:1} {65} {(4,1)} {id.} \hfill}
}
\hfill}
}
\hfill}

\vskip 10pt

As can be seen in the table, the same neighbouring weight with the same decomposition 
but a 
different current state may apply to different cells: as an example, 48 is the weight for entries 
34 and 35 sharing the neighbourhood {\footnotesize {20101}}.
As mentioned in the table, entry 34 applies to the central cell while entry 35 apply to the
cell (4,1). In figure~\ref{f_fix_m} entry 34 applies when the locomotive has its rear in the
cell (4,1) while entry 35 applies to that cell when the rear is in the cell (4,4).
\vskip 5pt
Let us now look at the converters. Table~\ref{t_convert} displays the entries of Table~\ref{t_small}
together with the cells to which those entries apply. The cells referred too can be seen on
Figure~\ref{f_ch_m}. We can see in lines (ch b) entry 54 whose neighbourhood is the maximal in
Table~\ref{t_small}. The neighbours are given by {\footnotesize{10032}} which indicates three
\MM-neighbours and two \VV-ones giving 155 from value 156 of the total neighbouring weight. Lines 
(ch b) and (ch m) refer to the central cell of Figure~\ref{f_ch_m}. In Table~\ref{t_convert}, we 
can see, entry 48, that the neighbourhood of cell~0 is {\footnotesize{11022}} when the front of a 
blue locomotive is seen by that cell. The \BB-front is the second \uu. The corresponding situation
in the opposite conversion is given by entry~63 giving {\footnotesize{12012}} as neighbourhood
fo cell~0. The \MM-front is the \uu{} at he penultimate digit.
\vskip 5pt
\ligne{\hfill
$\vcenter{\hbox{\vtop{\leftskip 0pt\parindent 0pt\hsize=250pt
\ligne{\hfill\hbox to \rclarge{\hfill time\hfill} 
\tttrc 0       1       2       3       4       5       6       7        {} \hfill}
\ligne{\hfill\hbox to \rclarge{\hfill state\hfill}
\tttrc {\YY}   {\YY}   {\YY}   {\MM}   {\RR}   {\YY}   {\YY}   {\YY}    {} \hfill}
\ligne{\hfill\hbox to \rclarge{\hfill sum\hfill}
\tttrc {128}    {128}    {131}    {139}    {156}    {128}    {128}    {128}     {} \hfill}
\ligne{\hfill\hbox to \rclarge{\hfill line\hfill}
\tttrc {45}    {45}    {48}    {50}    {54}    {58}    {45}    {45}     {} \hfill}
}}}$
\hfill (ch b)\hskip 15pt}
\vskip 10pt
\ligne{\hfill
$\vcenter{\hbox{\vtop{\leftskip 0pt\parindent 0pt\hsize=250pt
\ligne{\hfill\hbox to \rclarge{\hfill time\hfill} 
\tttrc 0       1       2       3       4       5       6       7        {} \hfill}
\ligne{\hfill\hbox to \rclarge{\hfill state\hfill}
\tttrc {\YY}   {\YY}   {\YY}   {\YY}   {\YY}   {\YY}   {\YY}   {\YY}    {} \hfill}
\ligne{\hfill\hbox to \rclarge{\hfill sum\hfill}
\tttrc {78}    {78}    {106}    {89}    {81}    {89}    {78}    {78}     {} \hfill}
\ligne{\hfill\hbox to \rclarge{\hfill line\hfill}
\tttrc {60}    {60}    {63}    {64}    {69}    {72}    {60}    {60}     {} \hfill}
}}}$
\hfill (ch m)\hskip 15pt}
\vskip 10pt
\ligne{\hfill
\vtop{\leftskip 0pt\parindent 0pt
\begin{tab}\label{t_convert}
\leurre
Entries of Table~\ref{t_small} devoted to the converters.
\end{tab}
\ligne{\hfill
\vtop{\leftskip 0pt\parindent 0pt\hsize=120pt
\ligne{\hfill\tttlcourt {n°} {{\footnotesize{c-YBRMV:n}}} {$\Sigma$} {c.} {F.} \hfill}
\vskip 3pt
\ligne{\hfill converters\hfill}
\ligne{\hfill blue to mauve\hfill}
\ligne{\hfill\tttlcourt { 45} {1-20022:1} {128} {0} {\ref{f_ch_m}} \hfill}
\ligne{\hfill\tttlcourt { 46} {4-10011:4} {64} {(1,1)} {id.} \hfill}
\ligne{\hfill\tttlcourt { 47} {5-20010:5} {31} {(3,1)} {id.} \hfill}
\ligne{\hfill\tttlcourt { 48} {1-11022:4} {131} {0} {id.}  \hfill}
\ligne{\hfill\tttlcourt { 49} {5-11010:5} {34} {(3,3)} {id.} \hfill}
\ligne{\hfill\tttlcourt { 50} {4-10122:3} {139} {0} {id.} \hfill}
\ligne{\hfill\tttlcourt { 51} {4-00021:4} {92} {(1,1)} {id.}  \hfill}
\ligne{\hfill\tttlcourt { 52} {3-10011:1} {64} {(6,1)} {id.} \hfill}
\ligne{\hfill\tttlcourt { 53} {5-10020:5} {59} {(3,1)} {id.}  \hfill}
\ligne{\hfill\tttlcourt { 54} {3-10032:1} {156} {0} {id.} \hfill}
\ligne{\hfill\tttlcourt { 55} {5-10012:5} {98} {(1,2)} {id.} \hfill}
\ligne{\hfill\tttlcourt { 56} {4-00111:4} {75} {(1,1)} {id.} \hfill}
\ligne{\hfill\tttlcourt { 57} {5-00120:5} {70} {(7,1)} {id.} \hfill}
\ligne{\hfill\tttlcourt { 58} {1-10122:1} {139} {0} {id.} \hfill}
\ligne{\hfill\tttlcourt { 59} {5-10110:5} {42} {(7,1)} {id.} \hfill}
}
\hfill
\vtop{\leftskip 0pt\parindent 0pt\hsize=120pt
\ligne{\hfill\tttlcourt {n°} {{\footnotesize{c-YBRMV:n}}} {$\Sigma$} {c.} {F.} \hfill}
\vskip 3pt
\ligne{\hfill mauve to blue\hfill}
\ligne{\hfill\tttlcourt { 60} {1-22002:1} {78} {0} {\ref{f_ch_m}} \hfill}
\ligne{\hfill\tttlcourt { 61} {5-21000:5} {6} {(3,1)} {id.} \hfill}
\ligne{\hfill\tttlcourt { 62} {2-11001:2} {39} {(1,1)} {id.} \hfill}
\ligne{\hfill\tttlcourt { 63} {1-12012:2} {106} {0} {id.} \hfill}
\ligne{\hfill\tttlcourt { 64} {2-12102:3} {89} {0} {id.} \hfill}
\ligne{\hfill\tttlcourt { 65} {5-02100:5} {20} {(7,1)} {id.} \hfill}
\ligne{\hfill\tttlcourt { 66} {5-12000:5} {9} {(3,1)} {id.} \hfill}
\ligne{\hfill\tttlcourt { 67} {2-02001:2} {42} {(1,1)} {id.} \hfill}
\ligne{\hfill\tttlcourt { 68} {3-11001:1} {39} {(6,1)} {id.}  \hfill}
\ligne{\hfill\tttlcourt { 69} {3-13002:1} {81} {0} {id.} \hfill}
\ligne{\hfill\tttlcourt { 70} {5-11100:5} {17} {(7,1)} {id.} \hfill}
\ligne{\hfill\tttlcourt { 71} {2-01101:2} {50} {(1,1)} {id.} \hfill}
\ligne{\hfill\tttlcourt { 72} {1-12102:1} {89} {0} {id.} \hfill}
}
\hfill}
}
\hfill}
\vskip 10pt
Table~\ref{t_filter} displays the entries of Table~\ref{t_small} devoted to the filters. Before, we 
display the entry which apply to the central cell together with the corresponding number of the
entry. Lines (ftb b) and (ftb m) show us the entries and the corresponding neighbouring weights
applied to cell~0 when a blue and a mauve locomotive respectively comes to the blue filter. 
Lines (ftm m) and (ftm b) fo the same for a mauve and a blue locomotive respectively with respect to
a mauve filter. In lines (ftb b) and (ftm m) we can see that the filter let a locomotive of the same
colour cross the filter. Lines (ftb m) and (ftm b) show us that the filter prevents the crossing
for a locomotive of an opposite colour. 
\vskip 10pt
\ligne{\hfill
$\vcenter{\hbox{\vtop{\leftskip 0pt\parindent 0pt\hsize=250pt
\ligne{\hfill\hbox to \rclarge{\hfill time\hfill} 
\tttrc 0       1       2       3       4       5       6       {}       {} \hfill}
\ligne{\hfill\hbox to \rclarge{\hfill state\hfill}
\tttrc {\YY}   {\YY}   {\BB}   {\RR}   {\YY}   {\YY}   {\YY}   {}    {} \hfill}
\ligne{\hfill\hbox to \rclarge{\hfill sum\hfill}
\tttrc {52}    {55}    {63}    {55}    {63}    {52}    {52}    {}     {} \hfill}
\ligne{\hfill\hbox to \rclarge{\hfill line\hfill}
\tttrc {73}    {84}    {91}    {97}    {102}   {73}    {73}    {}     {} \hfill}
}}}$
\hfill (ftb b)\hskip 15pt}
\vskip 10pt
\ligne{\hfill
$\vcenter{\hbox{\vtop{\leftskip 0pt\parindent 0pt\hsize=250pt
\ligne{\hfill\hbox to \rclarge{\hfill time\hfill} 
\tttrc 0       1       2       3       4       5       6       {}       {} \hfill}
\ligne{\hfill\hbox to \rclarge{\hfill state\hfill}
\tttrc {\YY}   {\YY}   {\YY}   {\YY}   {\YY}   {\YY}   {\YY}   {}    {} \hfill}
\ligne{\hfill\hbox to \rclarge{\hfill sum\hfill}
\tttrc {52}    {80}    {63}    {52}    {52}    {52}    {52}    {}     {} \hfill}
\ligne{\hfill\hbox to \rclarge{\hfill line\hfill}
\tttrc {73}    {105}    {102}    {73}    {73}    {73}    {73}    {}     {} \hfill}
}}}$
\hfill (ftb m)\hskip 15pt}

\vskip 10pt
\ligne{\hfill
$\vcenter{\hbox{\vtop{\leftskip 0pt\parindent 0pt\hsize=250pt
\ligne{\hfill\hbox to \rclarge{\hfill time\hfill} 
\tttrc 0       1       2       3       4       5       6       {}       {} \hfill}
\ligne{\hfill\hbox to \rclarge{\hfill state\hfill}
\tttrc {\YY}   {\YY}   {\MM}   {\RR}   {\YY}   {\YY}   {\YY}   {}    {} \hfill}
\ligne{\hfill\hbox to \rclarge{\hfill sum\hfill}
\tttrc {77}    {105}    {88}    {105}    {88}    {77}    {77}    {}     {} \hfill}
\ligne{\hfill\hbox to \rclarge{\hfill line\hfill}
\tttrc {111}    {118}    {119}    {124}    {127}    {111}    {111}    {}     {} \hfill}
}}}$
\hfill (ftm m)\hskip 15pt}
\vskip 10pt
\ligne{\hfill
$\vcenter{\hbox{\vtop{\leftskip 0pt\parindent 0pt\hsize=250pt
\ligne{\hfill\hbox to \rclarge{\hfill time\hfill} 
\tttrc 0       1       2       3       4       5       6       {}       {} \hfill}
\ligne{\hfill\hbox to \rclarge{\hfill state\hfill}
\tttrc {\YY}   {\YY}   {\YY}   {\YY}   {\YY}   {\YY}   {\YY}   {}    {} \hfill}
\ligne{\hfill\hbox to \rclarge{\hfill sum\hfill}
\tttrc {77}    {80}    {88}    {77}    {77}    {77}    {77}    {}     {} \hfill}
\ligne{\hfill\hbox to \rclarge{\hfill line\hfill}
\tttrc {111}    {105}    {127}    {111}    {111}    {111}    {111}    {}     {} \hfill}
}}}$
\hfill (ftm b)\hskip 15pt}
\vskip 10pt
In lines (ftb b) and (ftm m) we can see the sums 55, 63 and 105, 88 respectively appear twice
with different numbers of entries: 84 and 97 for sum 55, 63 and 102 for sum 91{} in (ftb b)
and also 118 and 124 for sum 105, 119 and 127 for sum 88{} in (ftm m). Note that entry~102, 127
is applied both for the blue, the mauve filter respectively, the second occurrence being for 
stopping the locomotive of the opposite colour.

Note that in Table~\ref{t_convert}, several entries are conservative or witnessing. As an example,
entries~73 up to~79 are conservative while entry~76, for instance, witnesses the front of a blue 
locomotive which is also the case, later for cell (1,1) with entry~92. In both cases, in the
neighbourhood witness of the states, the witnessing entry has less \YY-marks by one and one more
\BB-mark: we have {\footnotesize{31002}}, {\footnotesize{22002}} for entries 74, 85 respectively
and we have {\footnotesize{20020}}, {\footnotesize{11020}} for entries 77, 90 respectively.
\vskip 10pt
Table~\ref{t_prog} displays the last entries of Table~\ref{t_small} which are devoted to the
programming of the filters. Lines (ch m.f.) and (ch b.f.) show us which entries apply to the cell 
(1,1) of a filter, the cell which defines the colour of the filter and which is changed by a mauve 
locomotive arriving nearby it. We can see that two neighbouring sum occur: 163 and 164, both with 
two distinct entries: 129, 134 and 115, 75 respectively. Entries 129 and 134 change the colour of 
the filter: entry~129 changes a mauve filter to a blue one while entry~134 changes a blue filter to
a mauve one, as mentioned by the explanation of the entry, {\footnotesize{4-10003:2}} and
{\footnotesize{2-10003:4}} for entries 129 and 134 respectively. Note that the neighbourhoods are 
the same in both cases. Entries 75 and 115 are 
conservative: they keep the appropriate colour in the cell (1,1) which holds the colour of the 
filter. Entry 75 keeps a blue filter while entry 115 keeps a mauve one. The neighbours are the 
same in both cases in an idle configuration: {\footnotesize{20003}}.

\vskip 10pt
\ligne{\hfill
$\vcenter{\hbox{\vtop{\leftskip 0pt\parindent 0pt\hsize=250pt
\ligne{\hfill\hbox to \rclarge{\hfill time\hfill} 
\tttrc 0       1       2       3       4       5       6       7       {} \hfill}
\ligne{\hfill\hbox to \rclarge{\hfill state\hfill}
\tttrc {\MM}   {\MM}   {\BB}   {\BB}   {\BB}   {\BB}   {\BB}   {\BB}    {} \hfill}
\ligne{\hfill\hbox to \rclarge{\hfill sum\hfill}
\tttrc {104}   {103}   {104}   {104}   {104}   {104}   {104}   {104}    {} \hfill}
\ligne{\hfill\hbox to \rclarge{\hfill line\hfill}
\tttrc {115}    {129}    {75}   {75}   {75}   {75}    {75}    {75}     {} \hfill}
}}}$
\hfill (ch m.f.)\hskip 15pt}
\vskip 10pt 
\ligne{\hfill
$\vcenter{\hbox{\vtop{\leftskip 0pt\parindent 0pt\hsize=250pt
\ligne{\hfill\hbox to \rclarge{\hfill time\hfill} 
\tttrc 0       1       2       3       4       5       6       7       {} \hfill}
\ligne{\hfill\hbox to \rclarge{\hfill state\hfill}
\tttrc {\WW}   {\WW}   {\WW}   {\MM}   {\RR}   {\WW}   {\WW}   {\WW}    {} \hfill}
\ligne{\hfill\hbox to \rclarge{\hfill sum\hfill}
\tttrc {104}   {103}   {104}   {104}   {104}   {104}   {104}   {104}    {} \hfill}
\ligne{\hfill\hbox to \rclarge{\hfill line\hfill}
\tttrc {75}    {134}    {115}   {115}   {115}   {115}    {115}    {115}     {} \hfill}
}}}$
\hfill (ch b.f.)\hskip 15pt}
\vskip 10pt
\vskip 10pt
\ligne{\hfill
\vtop{\leftskip 0pt\parindent 0pt
\begin{tab}\label{t_filter}
\leurre
Entries of Table~\ref{t_small} devoted to the filters.
\end{tab}
\ligne{\hfill
\vtop{\leftskip 0pt\parindent 0pt\hsize=120pt
\ligne{\hfill\tttlcourt {n°} {{\footnotesize{c-YBRMV:n}}} {$\Sigma$} {c.} {F.} \hfill}
\vskip 3pt
\ligne{\hfill blue filter\hfill}
\ligne{\hfill permitting\hfill}
\ligne{\hfill\tttlcourt { 73} {1-21101:1} {52} {0} {\ref{f_sel_m}} \hfill}
\ligne{\hfill\tttlcourt { 74} {5-31002:5} {75} {(7,1)} {id.} \hfill}
\ligne{\hfill\tttlcourt { 75} {2-20003:2} {104} {(1,1)} {id.} \hfill}
\ligne{\hfill\tttlcourt { 76} {5-01001:5} {38} {(1,4)} {id.} \hfill}
\ligne{\hfill\tttlcourt { 77} {3-20020:3} {60} {(5,1)} {id.} \hfill}
\ligne{\hfill\tttlcourt { 78} {5-10002:5} {69} {(7,2)} {id.} \hfill}
\ligne{\hfill\tttlcourt { 79} {5-00002:5} {68} {(7,7)} {id.} \hfill}
\ligne{\hfill\tttlcourt { 80} {1-11102:2} {85} {(6,1)} {id.} \hfill}
\ligne{\hfill\tttlcourt { 81} {1-01002:1} {72} {(1,2)} {id.} \hfill}
\ligne{\hfill\tttlcourt { 82} {5-11001:5} {39} {(1,3)} {id.} \hfill}
\ligne{\hfill\tttlcourt { 83} {4-00100:4} {12} {(5,2)} {id.} \hfill}
\ligne{\hfill\tttlcourt { 84} {1-12101:2} {55} {0} {id.} \hfill}
\ligne{\hfill\tttlcourt { 85} {5-22002:5} {78} {(7,1)} {id.} \hfill}
\ligne{\hfill\tttlcourt { 86} {5-01002:5} {72} {(1,2)} {id.} \hfill}
\ligne{\hfill\tttlcourt { 87} {2-20002:2} {70} {(7,4)} {\ref{f_flt_m}} \hfill}
\ligne{\hfill\tttlcourt { 88} {5-11000:5} {5} {(6,8)} {\ref{f_sel_m}} \hfill}
\ligne{\hfill\tttlcourt { 89} {2-10202:3} {93} {(6,1)} {id.} \hfill}
\ligne{\hfill\tttlcourt { 90} {3-11020:3} {63} {(5,1)} {id.} \hfill}
\ligne{\hfill\tttlcourt { 91} {2-11201:3} {63} {0} {id.} \hfill}
\ligne{\hfill\tttlcourt { 92} {2-11003:2} {107} {(1,1)} {id.} \hfill}
\ligne{\hfill\tttlcourt { 93} {5-12102:5} {89} {(7,1)} {id.} \hfill}
\ligne{\hfill\tttlcourt { 94} {5-00102:5} {80} {0} {id.} \hfill}
\ligne{\hfill\tttlcourt { 95} {3-11102:1} {85} {(6,1)} {id.} \hfill}
\ligne{\hfill\tttlcourt { 96} {3-01120:3} {74} {(5,1)} {id.} \hfill}
\ligne{\hfill\tttlcourt { 97} {3-12101:1} {55} {0} {id.} \hfill}
\ligne{\hfill\tttlcourt { 98} {2-10103:2} {115} {(1,1)} {id.} \hfill}
\ligne{\hfill\tttlcourt { 99} {5-21102:5} {86} {(7,1)} {id.} \hfill}
\ligne{\hfill\tttlcourt {100} {1-10202:1} {93} {(6,1)} {id.} \hfill}
}
\hfill
\vtop{\leftskip 0pt\parindent 0pt\hsize=120pt
\ligne{\hfill\tttlcourt {n°} {{\footnotesize{c-YBRMV:n}}} {$\Sigma$} {c.} {F.} \hfill}
\vskip 3pt
\ligne{\hfill\tttlcourt {101} {3-10120:3} {71} {(5,1)} {id.} \hfill}
\ligne{\hfill\tttlcourt {102} {1-11201:1} {63} {0} {\ref{f_ch_m}} \hfill}
\ligne{\hfill\tttlcourt {103} {1-20102:1} {82} {(6,1)} {id.} \hfill}
\ligne{\hfill stopping\hfill}
\ligne{\hfill\tttlcourt {104} {1-10112:4} {110} {(6,1)} {id.} \hfill}
\ligne{\hfill\tttlcourt {105} {1-11111:1} {80} {0} {id.} \hfill}
\ligne{\hfill\tttlcourt {106} {5-21012:5} {103} {(7,1)} {id.} \hfill}
\ligne{\hfill\tttlcourt {107} {5-00012:5} {97} {(1,2)} {id.} \hfill}
\ligne{\hfill\tttlcourt {108} {4-10202:3} {93} {(6,1)} {id.} \hfill}
\ligne{\hfill\tttlcourt {109} {3-10030:3} {88} {(5,1)} {id.} \hfill}
\ligne{\hfill\tttlcourt {110} {3-20102:1} {82} {(6,1)} {id.} \hfill}
\ligne{\hfill mauve filter\hfill}
\ligne{\hfill\tttlcourt {111} {1-20111:1} {77} {0} {id.} \hfill}
\ligne{\hfill\tttlcourt {112} {1-00012:1} {97} {(1,2)} {id.} \hfill}
\ligne{\hfill\tttlcourt {113} {5-10011:5} {64} {(1,3)} {id.} \hfill}
\ligne{\hfill\tttlcourt {114} {5-00011:5} {63} {(1,4)} {id.} \hfill}
\ligne{\hfill\tttlcourt {115} {4-20003:4} {104} {(1,1)} {id.} \hfill}
\ligne{\hfill\tttlcourt {116} {5-20022:5} {128} {(7,1)} {id.} \hfill}
\ligne{\hfill\tttlcourt {117} {5-30012:5} {100} {(7,1)} {id.} \hfill}
\ligne{\hfill\tttlcourt {118} {1-10121:4} {105} {0} {id.} \hfill}
\ligne{\hfill\tttlcourt {119} {4-10211:3} {88} {0} {id.} \hfill}
\ligne{\hfill\tttlcourt {120} {5-10122:5} {139} {(7,1)} {id.} \hfill}
\ligne{\hfill\tttlcourt {121} {4-10013:4} {132} {(1,1)} {id.} \hfill}
\ligne{\hfill\tttlcourt {122} {3-10112:1} {110} {(6,1)} {id.} \hfill}
\ligne{\hfill\tttlcourt {123} {3-00130:3} {99} {(5,1)} {id.} \hfill}
\ligne{\hfill\tttlcourt {124} {3-10121:1} {105} {0} {id.} \hfill}
\ligne{\hfill\tttlcourt {125} {4-10103:4} {115} {(1,1)} {id.} \hfill}
\ligne{\hfill\tttlcourt {126} {5-20112:5} {111} {(7,1)} {id.} \hfill}
\ligne{\hfill\tttlcourt {127} {1-10211:1} {88} {0} {id.} \hfill}
}
\hfill}
}
\hfill}

\ligne{\hfill
\vtop{\leftskip 0pt\parindent 0pt
\begin{tab}\label{t_prog}
\leurre
The table displays the entries of Table~\ref{t_small} devoted to the control of the filters.
\end{tab}
\ligne{\hfill
\vtop{\leftskip 0pt\parindent 0pt\hsize=120pt
\ligne{\hfill\tttlcourt {n°} {{\footnotesize{c-YBRMV:n}}} {$\Sigma$} {c.} {F.} \hfill}
\vskip 3pt
\ligne{\hfill change filters\hfill}
\ligne{\hfill mauve to blue\hfill}
\ligne{\hfill\tttlcourt {128} {1-10012:1} {98} {(1,2)} {\ref{f_flt_m}} \hfill}
\ligne{\hfill\tttlcourt {129} {4-10003:2} {103} {(1,1)} {id.} \hfill}
\ligne{\hfill\tttlcourt {130} {5-20012:5} {99} {(7,1)} {id.} \hfill}
\ligne{\hfill\tttlcourt {131} {0-00112:1} {109} {(1,2)} {id.} \hfill}
}
\hfill
\vtop{\leftskip 0pt\parindent 0pt\hsize=120pt
\ligne{\hfill\tttlcourt {n°} {{\footnotesize{c-YBRMV:n}}} {$\Sigma$} {c.} {F.} \hfill}
\vskip 3pt
\ligne{\hfill\tttlcourt {132} {3-10000:1} {1} {(5,1)} {\ref{f_flt_m}} \hfill}
\ligne{\hfill\tttlcourt {133} {1-11002:1} {73} {(1,2)} {id.} \hfill}
\ligne{\hfill blue to mauve\hfill}
\ligne{\hfill\tttlcourt {134} {2-10003:4} {103} {(1,1)} {id.} \hfill}
\ligne{\hfill\tttlcourt {135} {0-01102:1} {84} {(1,2)} {id.} \hfill}
\ligne{\hfill\tttlcourt {136} {5-21002:5} {74} {(5,1)} {id.} \hfill}
}
\hfill}
}
\hfill}

\ifnum 1=0 {
Line (chbl) applies to a mauve filter which stops blue locomotives. Accordingly the neighbourhood
of the cell~(4,2) contains two \MM-cells together with a \GG-one together with a \YY-one. That
justifies the weight 216. That weight becomes 228 when (4,2) can see the front of the mauve 
locomotive which is the signal for changing the colour of the filter. Which occurs when the rear
of the locomotive is in (4,2). Indeed, after that, the weight is 194 which corresponds to the \GG- 
and \YY-neighbours and the two \BB-neighbours indicating the new colour of the filter. Lines (chok)
indicates the lines for the opposite operation which also is clear from the weights.

Note that we already met lines 159 and 154{} in the context of the fork. Let us look at the case of
line 154. Table~\ref{t_prog} indicates one \GG-, one \YY- and two \MM-neighbours. Note on 
Figure~\ref{f_auxil} that the positions of those neighbours in the case of the cell~0 of the fork 
and in the case of the cell~(4,2) of a mauve filter are different. In the case of the filter, the
two \MM-cells are also neighbours of one another which is not the case in the configuration of the
fork, see Figure~\ref{f_frk_m}. A similar comment applies to line~159 for cell~0 of the fork and 
for the cell~(4,2) of the filter when, in the case of the fork, the locomotive is blue while in the
case of the filter, the signal deals with a blue filter, transforming if into a mauve one.
} \fi
\vskip 10pt
As far as we had a look on Tables~\ref{t_tracks} up to \ref{t_prog}, scrutinising them, we can see
that we have the following maximal coefficients of the weights to the states in the computation of
the neighbourhood weight:
\newdimen\wwlarge\wwlarge=25pt
\def\tttww #1 #2 #3 #4 #5 #6 {%
\hbox to \wwlarge {\hfill#1\hfill}
\hbox to \wwlarge {\hfill#2\hfill}
\hbox to \wwlarge {\hfill#3\hfill}
\hbox to \wwlarge {\hfill#4\hfill}
\hbox to \wwlarge {\hfill#5\hfill}
\hbox to \wwlarge {\hfill#6\hfill}
}
\vskip 5pt
\slarge=35pt
\ligne{\hfill
$\vcenter{\hbox{\vtop{\leftskip 0pt\parindent 0pt\hsize=300pt
\ligne{\hfill \hbox to \slarge {state\hfill} 
\tttww {\WW} {\YY} {\BB} {\RR} {\MM} {\VV} \hfill} 
\ligne{\hfill \hbox to \slarge {rank\hfill} 
\tttww 0 1 2 3 4 5 \hfill} 
\ligne{\hfill \hbox to \slarge {weight\hfill} 
\tttww {0}  {1}   {4}   {12}  {29}  {34} \hfill}
\ligne{\hfill \hbox to \slarge {coeff.\hfill} 
\tttww {$x$}   {3}   {3}   {2}   {3}   {3} \hfill}
}}}$\hfill(\numerrel)\hskip 15pt}
\vskip 10pt
\noindent
where in (6), $x$ means that the coefficient for \WW{} in any entry is the complement to 7 of the 
sum of the coefficients of the other states.

\slarge=25pt
Let $w_i$ with \hbox{$i\in \{0..5\}$} be the weight given to the state of rank $i$. Let $\kappa_i$
be the maximal coefficient for $w_i$ in Tables~\ref{t_tracks} up to \ref{t_prog}. If we put
\vskip 5pt
\ligne{\hfill
$w_i = 1+\displaystyle{\sum_{0\leq j<i}\kappa_j . w_j}$
\hfill(\numerrel)\hskip 15pt}

\noindent
we get a sufficient condition for obtaining the uniqueness of decomposing any $n\leq \kappa_4.w_4$
as
%If we fix $w_0=0$ and $w_1=1$, weights satisfying (7) are given by the weights $w_i$ displayed in 
%(6) for $i<5$. We thus get that $w_4 = 49$ where $w_4$ is the weight of \MM. 
%It is not difficult to show that any positive number $n$ with $n\leq \kappa_4.w_4$ can be 
%uniquely represented as 
\vskip 5pt
\ligne{\hfill
$n=\displaystyle{\sum_{0\leq j<4}\alpha_j . w_j}$ \hfill (\numerrel)\hskip 15pt}
\vskip 5pt
\noindent
with $\alpha_j\leq\kappa_j$ for $j\in\{0..4\}$ by taking $w_0=0$ and $w_1=1$.
\ifnum 1=0 {
It is also not difficult to prove that from (7) %and from $w_5<w_6$ 
we have that
\vskip 5pt
\ligne{\hfill
$\vcenter{\hbox{\vtop{\leftskip 0pt\parindent 0pt\hsize=260pt
$\displaystyle{\sum_{i\leq 5}\alpha_i.w_i}=\displaystyle{\sum_{i\leq 5}\beta_i.w_i}$ with 
$\alpha_i,\beta_i\leq\kappa_i$ for $i\in\{1..5\}$\\ $\Rightarrow$ $\alpha_i=\beta_i$ for 
$i \in\{1..5\}$.}}}$
\hfill(\numerrel)\hskip 15pt}
\vskip 5pt
} \fi

Not any $n$ is reached by a sum as in (8) by the decomposition shown in
Tables~\ref{t_tracks} up to~\ref{t_prog}. On the 137 entries, there are 
69 pairwise distinct values. Accordingly many neighbouring weights are duplicated, always associated
with a different state. Nonetheless, it is worth to set $w_5 > w_4$ where $w_5$ is the weight 
of ~\VV. Note that 34 is reached by entries, 22 and 49, with the same state but with a
different decomposition: {\footnotesize {00001}} and {\footnotesize {11010}} for entries 22 and 49
respectively. In Table~\ref{t_small}, the new state is uniquely defined by the old state and 
the neighbourhood weight. It is another reason to set $w_5>w_4$. The values satisfying (7) are,
from 0 to~4 : 1, 4, 14 and 30. Note that in Table~\ref{t_small}, there is at most 2 occurrences of
\RR{} if there is at least 1 occurrence of~\MM. So that it is enough to take 12 for \RR{} and then
29 for \MM. Starting from 30, the first value of $w_5> w_4$ which makes correct pictures is 34.
For smaller values of $w_2$ or $w_4$, there are incorrect pictures. Accordingly, as far as those 
pictures are conformal to what is explained in Subsection~\ref{newrailway}, Theorem~\ref{letheo} is 
thus completely proved. \hfill $\Box$

\ifnum 1=0 {
\newdimen\llfilt\llfilt=25pt
\def\tttfilt #1 #2 #3 #4 #5 #6 #7 #8 {%
\hbox to \llfilt {\hfill#1\hfill}
\hbox to \llfilt {\hfill#2\hfill}
\hbox to \llfilt {\hfill#3\hfill}
\hbox to \llfilt {\hfill#4\hfill}
\hbox to \llfilt {\hfill#5\hfill}
\hbox to \llfilt {\hfill#6\hfill}
\hbox to \llfilt {\hfill#7\hfill}
\hbox to \llfilt {\hfill#8\hfill}
}

\ligne{\hfill
\vtop{\leftskip 0pt\parindent 0pt\hsize=220pt
\begin{tab}\label{t_sample}
\leurre
The transformation of an opening filter into a closing one.
\end{tab}
\ligne{\hfill\tttfilt 0     {3,1)} {(4,1)} {(4,2)} {(4,5)} {(4,6)} {(4,7)} {(4,18)} \hfill}
\ligne{\hfill\tttfilt {\VV} {\MM}  {\MM}   {\WW}   {\YY}   {\WW}   {\GG}   {\MM} \hfill}
\ligne{\hfill\tttfilt {106} {103}  {103}   {d.}    {13}    {39}    {31}    {40}  \hfill}
\ligne{\hfill\tttfilt {106} {103}  {103}   {152}   {37}    {40}    {31}    {41}  \hfill}
\ligne{\hfill\tttfilt {106} {154}  {154}   {153}   {38}    {41}    {31}    {d.} \hfill}
\ligne{\hfill\tttfilt {106} {155}  {155}   {156}   {15}    {d.}    {8}     {17}  \hfill}
\ligne{\hfill\tttfilt {66}  {63}   {63}    {d.}    {13}    {18}    {48}    {17}  \hfill}
}
\hfill}
\vskip 10pt
} \fi

\ligne{\hfill
\vtop{
\begin{tab}\label{t_small}
\leurre
The table for the transition function. In the column 'n°', the number of the instruction;
{\footnotesize{c:YBRMV:n}} refers to the current state, the number of neighbours in the states
\YY, \BB, \RR, \MM{} or \VV, and $\Sigma$ refers to the neighbourhood weight.
\end{tab}
%\vspace{-10pt}
%
\ligne{\hfill
\vtop{\leftskip 0pt\parindent 0pt\hsize=85pt
\ligne{\hfill\tttncourt {n°} {\footnotesize{c:YBRMV:n}} {$\Sigma$} {} \hfill}
\vskip 3pt
\ligne{\hrulefill\ \ tracks\ \ \hrulefill}
\ligne{\hfill\tttncourt {  0} {0-00000:0} {0} {} \hfill}
\ligne{\hfill\tttncourt {  1} {1-20000:1} {2} {} \hfill}
\ligne{\hfill\tttncourt {  2} {1-10000:1} {1} {} \hfill}
\ligne{\hfill blue locomotive}
\ligne{\hfill\tttncourt {  3} {1-11000:2} {5} {} \hfill}
\ligne{\hfill\tttncourt {  4} {2-10100:3} {13} {} \hfill}
\ligne{\hfill\tttncourt {  5} {3-11000:1} {5} {} \hfill}
\ligne{\hfill\tttncourt {  6} {1-10100:1} {13} {} \hfill}
\ligne{\hfill\tttncourt {  7} {1-00100:1} {12} {} \hfill}
\ligne{\hfill mauve locomotive}
\ligne{\hfill\tttncourt {  8} {1-10010:4} {30} {} \hfill}
\ligne{\hfill\tttncourt {  9} {4-10100:3} {13} {} \hfill}
\ligne{\hfill\tttncourt { 10} {3-10010:1} {30} {} \hfill}
\ligne{\hrulefill\ \ fork\ \ \hrulefill}
\ligne{\hfill\tttncourt { 11} {1-30000:1} {3} {} \hfill}
\ligne{\hfill blue locomotive}
\ligne{\hfill\tttncourt { 12} {1-21000:2} {6} {} \hfill}
\ligne{\hfill\tttncourt { 13} {2-20100:3} {14} {} \hfill}
\ligne{\hfill\tttncourt { 14} {3-12000:1} {9} {} \hfill}
\ligne{\hfill\tttncourt { 15} {1-10200:1} {25} {} \hfill}
\ligne{\hfill mauve locomotive}
\ligne{\hfill\tttncourt { 16} {1-20010:4} {31} {} \hfill}
\ligne{\hfill\tttncourt { 17} {4-20100:3} {14} {} \hfill}
\ligne{\hfill\tttncourt { 18} {3-10020:1} {59} {} \hfill}
\ligne{\hrulefill\ \ fixed switch\ \ \hrulefill}
\ligne{\hfill\tttncourt { 19} {1-30001:1} {37} {} \hfill}
\ligne{\hfill\tttncourt { 20} {1-20001:1} {36} {} \hfill}
\ligne{\hfill\tttncourt { 21} {5-30001:5} {37} {} \hfill}
\ligne{\hfill\tttncourt { 22} {5-00001:5} {34} {} \hfill}
\ligne{\hfill blue locomotive}
\ligne{\hfill\tttncourt { 23} {1-11001:2} {39} {} \hfill}
\ligne{\hfill\tttncourt { 24} {2-10101:3} {47} {} \hfill}
\ligne{\hfill\tttncourt { 25} {1-21001:2} {40} {} \hfill}
\ligne{\hfill\tttncourt { 26} {5-21001:3} {40} {} \hfill}
\ligne{\hfill\tttncourt { 27} {3-11100:1} {17} {} \hfill}
\ligne{\hfill\tttncourt { 28} {1-11100:1} {17} {} \hfill}
\ligne{\hfill\tttncourt { 29} {2-20200:3} {26} {} \hfill}
\ligne{\hfill\tttncourt { 30} {3-11101:5} {51} {} \hfill}
\ligne{\hfill\tttncourt { 31} {5-00100:5} {12} {} \hfill}
}
\hfill
\vtop{\leftskip 0pt\parindent 0pt\hsize=85pt
\ligne{\hfill\tttncourt {n°} {\footnotesize{c:YBRMV:n}} {$\Sigma$} {} \hfill}
\vskip 3pt
\ligne{\hfill\tttncourt { 32} {3-21001:1} {40} {} \hfill}
\ligne{\hfill\tttncourt { 33} {1-10101:1} {47} {} \hfill}
\ligne{\hfill\tttncourt { 34} {5-20101:5} {48} {} \hfill}
\ligne{\hfill\tttncourt { 35} {1-20101:1} {48} {} \hfill}
\ligne{\hfill mauve locomotive}
\ligne{\hfill\tttncourt { 36} {1-10011:4} {64} {} \hfill}
\ligne{\hfill\tttncourt { 37} {4-10101:3} {47} {} \hfill}
\ligne{\hfill\tttncourt { 38} {1-20011:4} {65} {} \hfill}
\ligne{\hfill\tttncourt { 39} {5-20011:3} {65} {} \hfill}
\ligne{\hfill\tttncourt { 40} {3-10110:1} {42} {} \hfill}
\ligne{\hfill\tttncourt { 41} {1-10110:1} {42} {} \hfill}
\ligne{\hfill\tttncourt { 42} {4-20200:3} {26} {} \hfill}
\ligne{\hfill\tttncourt { 43} {3-10111:5} {76} {} \hfill}
\ligne{\hfill\tttncourt { 44} {3-20011:1} {65} {} \hfill}
\ligne{\hrulefill\ \ converters\ \ \hrulefill}
\ligne{\hfill blue to mauve\hfill}
\ligne{\hfill\tttncourt { 45} {1-20022:1} {128} {} \hfill}
\ligne{\hfill\tttncourt { 46} {4-10011:4} {64} {} \hfill}
\ligne{\hfill\tttncourt { 47} {5-20010:5} {31} {} \hfill}
\ligne{\hfill\tttncourt { 48} {1-11022:4} {131} {} \hfill}
\ligne{\hfill\tttncourt { 49} {5-11010:5} {34} {} \hfill}
\ligne{\hfill\tttncourt { 50} {4-10122:3} {139} {} \hfill}
\ligne{\hfill\tttncourt { 51} {4-00021:4} {92} {} \hfill}
\ligne{\hfill\tttncourt { 52} {3-10011:1} {64} {} \hfill}
\ligne{\hfill\tttncourt { 53} {5-10020:5} {59} {} \hfill}
\ligne{\hfill\tttncourt { 54} {3-10032:1} {156} {} \hfill}
\ligne{\hfill\tttncourt { 55} {5-10012:5} {98} {} \hfill}
\ligne{\hfill\tttncourt { 56} {4-00111:4} {75} {} \hfill}
\ligne{\hfill\tttncourt { 57} {5-00120:5} {70} {} \hfill}
\ligne{\hfill\tttncourt { 58} {1-10122:1} {139} {} \hfill}
\ligne{\hfill\tttncourt { 59} {5-10110:5} {42} {} \hfill}
\ligne{\hfill mauve to blue\hfill}
\ligne{\hfill\tttncourt { 60} {1-22002:1} {78} {} \hfill}
\ligne{\hfill\tttncourt { 61} {5-21000:5} {6} {} \hfill}
\ligne{\hfill\tttncourt { 62} {2-11001:2} {39} {} \hfill}
\ligne{\hfill\tttncourt { 63} {1-12012:2} {106} {} \hfill}
\ligne{\hfill\tttncourt { 64} {2-12102:3} {89} {} \hfill}
\ligne{\hfill\tttncourt { 65} {5-02100:5} {20} {} \hfill}
\ligne{\hfill\tttncourt { 66} {5-12000:5} {9} {} \hfill}
}
\hfill
\vtop{\leftskip 0pt\parindent 0pt\hsize=85pt
\ligne{\hfill\tttncourt {n°} {\footnotesize{c:YBRMV:n}} {$\Sigma$} {} \hfill}
\vskip 3pt
\ligne{\hfill\tttncourt { 67} {2-02001:2} {42} {} \hfill}
\ligne{\hfill\tttncourt { 68} {3-11001:1} {39} {} \hfill}
\ligne{\hfill\tttncourt { 69} {3-13002:1} {81} {} \hfill}
\ligne{\hfill\tttncourt { 70} {5-11100:5} {17} {} \hfill}
\ligne{\hfill\tttncourt { 71} {2-01101:2} {50} {} \hfill}
\ligne{\hfill\tttncourt { 72} {1-12102:1} {89} {} \hfill}
\ligne{\hrulefill\ \ blue filter\ \ \hrulefill}
\ligne{\hfill permitting\hfill}
\ligne{\hfill\tttncourt { 73} {1-21101:1} {52} {} \hfill}
\ligne{\hfill\tttncourt { 74} {5-31002:5} {75} {} \hfill}
\ligne{\hfill\tttncourt { 75} {2-20003:2} {104} {} \hfill}
\ligne{\hfill\tttncourt { 76} {5-01001:5} {38} {} \hfill}
\ligne{\hfill\tttncourt { 77} {3-20020:3} {60} {} \hfill}
\ligne{\hfill\tttncourt { 78} {5-10002:5} {69} {} \hfill}
\ligne{\hfill\tttncourt { 79} {5-00002:5} {68} {} \hfill}
\ligne{\hfill\tttncourt { 80} {1-11102:2} {85} {} \hfill}
\ligne{\hfill\tttncourt { 81} {1-01002:1} {72} {} \hfill}
\ligne{\hfill\tttncourt { 82} {5-11001:5} {39} {} \hfill}
\ligne{\hfill\tttncourt { 83} {4-00100:4} {12} {} \hfill}
\ligne{\hfill\tttncourt { 84} {1-12101:2} {55} {} \hfill}
\ligne{\hfill\tttncourt { 85} {5-22002:5} {78} {} \hfill}
\ligne{\hfill\tttncourt { 86} {5-01002:5} {72} {} \hfill}
\ligne{\hfill\tttncourt { 87} {2-20002:2} {70} {} \hfill}
\ligne{\hfill\tttncourt { 88} {5-11000:5} {5} {} \hfill}
\ligne{\hfill\tttncourt { 89} {2-10202:3} {93} {} \hfill}
\ligne{\hfill\tttncourt { 90} {3-11020:3} {63} {} \hfill}
\ligne{\hfill\tttncourt { 91} {2-11201:3} {63} {} \hfill}
\ligne{\hfill\tttncourt { 92} {2-11003:2} {107} {} \hfill}
\ligne{\hfill\tttncourt { 93} {5-12102:5} {89} {} \hfill}
\ligne{\hfill\tttncourt { 94} {5-00102:5} {80} {} \hfill}
\ligne{\hfill\tttncourt { 95} {3-11102:1} {85} {} \hfill}
\ligne{\hfill\tttncourt { 96} {3-01120:3} {74} {} \hfill}
\ligne{\hfill\tttncourt { 97} {3-12101:1} {55} {} \hfill}
\ligne{\hfill\tttncourt { 98} {2-10103:2} {115} {} \hfill}
\ligne{\hfill\tttncourt { 99} {5-21102:5} {86} {} \hfill}
\ligne{\hfill\tttncourt {100} {1-10202:1} {93} {} \hfill}
\ligne{\hfill\tttncourt {101} {3-10120:3} {71} {} \hfill}
\ligne{\hfill\tttncourt {102} {1-11201:1} {63} {} \hfill}
\ligne{\hfill\tttncourt {103} {1-20102:1} {82} {} \hfill}
}
\hfill}
}
\hfill}

\ligne{\hfill
\vtop{\leftskip 0pt\parindent 0pt\hsize=85pt
\ligne{\hfill\tttncourt {n°} {\footnotesize{c:YBRMV:n}} {$\Sigma$} {} \hfill}
\vskip 3pt
\ligne{\hfill stopping\hfill}
\ligne{\hfill\tttncourt {104} {1-10112:4} {110} {} \hfill}
\ligne{\hfill\tttncourt {105} {1-11111:1} {80} {} \hfill}
\ligne{\hfill\tttncourt {106} {5-21012:5} {103} {} \hfill}
\ligne{\hfill\tttncourt {107} {5-00012:5} {97} {} \hfill}
\ligne{\hfill\tttncourt {108} {4-10202:3} {93} {} \hfill}
\ligne{\hfill\tttncourt {109} {3-10030:3} {88} {} \hfill}
\ligne{\hfill\tttncourt {110} {3-20102:1} {82} {} \hfill}
\ligne{\hrulefill\ \ mauve filter\ \ \hrulefill}
\ligne{\hfill\tttncourt {111} {1-20111:1} {77} {} \hfill}
\ligne{\hfill\tttncourt {112} {1-00012:1} {97} {} \hfill}
\ligne{\hfill\tttncourt {113} {5-10011:5} {64} {} \hfill}
\ligne{\hfill\tttncourt {114} {5-00011:5} {63} {} \hfill}
}
\hfill
\vtop{\leftskip 0pt\parindent 0pt\hsize=85pt
\ligne{\hfill\tttncourt {n°} {\footnotesize{c:YBRMV:n}} {$\Sigma$} {} \hfill}
\vskip 3pt
\ligne{\hfill\tttncourt {115} {4-20003:4} {104} {} \hfill}
\ligne{\hfill\tttncourt {116} {5-20022:5} {128} {} \hfill}
\ligne{\hfill\tttncourt {117} {5-30012:5} {100} {} \hfill}
\ligne{\hfill\tttncourt {118} {1-10121:4} {105} {} \hfill}
\ligne{\hfill\tttncourt {119} {4-10211:3} {88} {} \hfill}
\ligne{\hfill\tttncourt {120} {5-10122:5} {139} {} \hfill}
\ligne{\hfill\tttncourt {121} {4-10013:4} {132} {} \hfill}
\ligne{\hfill\tttncourt {122} {3-10112:1} {110} {} \hfill}
\ligne{\hfill\tttncourt {123} {3-00130:3} {99} {} \hfill}
\ligne{\hfill\tttncourt {124} {3-10121:1} {105} {} \hfill}
\ligne{\hfill\tttncourt {125} {4-10103:4} {115} {} \hfill}
\ligne{\hfill\tttncourt {126} {5-20112:5} {111} {} \hfill}
\ligne{\hfill\tttncourt {127} {1-10211:1} {88} {} \hfill}
}
\hfill
\vtop{\leftskip 0pt\parindent 0pt\hsize=85pt
\ligne{\hfill\tttncourt {n°} {\footnotesize{c:YBRMV:n}} {$\Sigma$} {} \hfill}
\vskip 3pt
\ligne{\hrulefill\ \ change filters\ \ \hrulefill}
\ligne{\hfill mauve to blue\hfill}
\ligne{\hfill\tttncourt {128} {1-10012:1} {98} {} \hfill}
\ligne{\hfill\tttncourt {129} {4-10003:2} {103} {} \hfill}
\ligne{\hfill\tttncourt {130} {5-20012:5} {99} {} \hfill}
\ligne{\hfill\tttncourt {131} {0-00112:1} {109} {} \hfill}
\ligne{\hfill\tttncourt {132} {3-10000:1} {1} {} \hfill}
\ligne{\hfill\tttncourt {133} {1-11002:1} {73} {} \hfill}
\ligne{\hfill blue to mauve\hfill}
\ligne{\hfill\tttncourt {134} {2-10003:4} {103} {} \hfill}
\ligne{\hfill\tttncourt {135} {0-01102:1} {84} {} \hfill}
\ligne{\hfill\tttncourt {136} {5-21002:5} {74} {} \hfill}
}
\hfill}

\section{Conclusion}

    As I already discussed the point dealing with states and weights, I think it is now worth to 
compare the present paper with previous ones. 

    In the present paper, I use the same model of the simulation of a register machine by a
locomotive running over an appropriate circuit. However, the implementation of the circuit is
different in many regards. The way the two types of locomotives is implemented allowed me to go back
to an almost direct implementation of crossings. Another important point is the use of the filters.
As far as the filters are directly constructed as programmable allowed me to greatly simplify the
conception of various structures allowing me to implement the model. Accordingly, we get a more
economic solution in terms of number of cells involved in structures and also a more efficient one. 
The price to pay is that the diagrams are a bit less readable. May be another model could be 
explored.

    And so, there are a lot of open questions.

\end{document}